\documentclass{aa}  

\usepackage{graphicx}
\usepackage{txfonts}
\usepackage{color}
\usepackage{hyperref}
\usepackage{newtxtext,newtxmath}
\usepackage{amsmath}	

\DeclareRobustCommand{\VAN}[3]{#2}
\let\VANthebibliography\thebibliography
\def\thebibliography{\DeclareRobustCommand{\VAN}[3]{##3}\VANthebibliography}

\newcommand{\mrm}[1]{\mathrm{#1}}
\newcommand{\nuc}[2]{$\mrm{^{#2}#1}$}

\usepackage[normalem]{ulem}

\begin{document}

   \title{Gamma-ray line emission from the Local Bubble}


   \author{Thomas Siegert\inst{\ref{inst:jmu}}\thanks{email: thomas.siegert@uni-wuerzburg.de}
          \and
          Michael M. Schulreich\inst{\ref{inst:tub}}
          \and
          Niklas Bauer\inst{\ref{inst:jmu}}
          \and
          Rudi Reinhardt\inst{\ref{inst:jmu}}
          \and
          Saurabh Mittal\inst{\ref{inst:jmu}}
          \and
          Hiroki Yoneda\inst{\ref{inst:jmu},\ref{inst:riken}}
          }

   \institute{Julius-Maximilians-Universität Würzburg, Fakultät für Physik und Astronomie, Institut für Theoretische Physik und Astrophysik, Lehrstuhl für Astronomie, Emil-Fischer-Str.~31, D-97074 Würzburg, Germany
   \label{inst:jmu}\\
   \email{thomas.siegert@uni-wuerzburg.de}
   \and
   Zentrum für Astronomie und Astrophysik, Technische Universität Berlin, Hardenbergstr.~36, D-10623 Berlin, Germany
   \label{inst:tub}
   \and
   RIKEN Nishina Center, 2-1 Hirosawa, Wako, Saitama 351-0198, Japan
   \label{inst:riken}
   }

   \date{Received April 11, 2024; accepted XXX XX XXXX}

 
  \abstract
    %
    {
    Deep-sea archives that include intermediate-lived radioactive \nuc{Fe}{60} particles suggest the occurrence of several recent supernovae inside the present-day volume of the Local Bubble during the last $\sim$10\,Myr.
    The isotope \nuc{Fe}{60} is mainly produced in massive stars and ejected in supernova explosions, which should always result in a sizeable yield of \nuc{Al}{26} from the same objects.
    \nuc{Fe}{60} and \nuc{Al}{26} decay with lifetimes of 3.82 and 1.05\,Myr, and emit $\gamma$-rays at 1332 and 1809\,keV, respectively.
    These $\gamma$-rays have been measured as diffuse glow of the Milky Way, and would also be expected from inside the Local Bubble as foreground emission.
    Based on two scenarios, one employing a geometrical model and the other state-of-the-art hydrodynamics simulations, we estimate the expected fluxes of the 1332 and 1809\,keV $\gamma$-ray lines, as well as the resulting 511\,keV line from positron annihilation due to the \nuc{Al}{26} $\beta^+$-decay.
    We find fluxes in the range of $10^{-6}$--$10^{-5}\,\mathrm{ph\,cm^{-2}\,s^{-1}}$ for all three lines with isotropic contributions of 10--50\,\%.
    We show that these fluxes are within reach for the upcoming COSI-SMEX $\gamma$-ray telescope over its nominal satellite mission duration of 2\,yr.
    Given the Local Bubble models considered, we conclude that in the case of 10--20\,Myr-old superbubbles, the distributions of \nuc{Fe}{60} and \nuc{Al}{26} are not co-spatial -- an assumption usually made in $\gamma$-ray data analyses.
    In fact, this should be taken into account however when analysing individual nearby targets for their \nuc{Fe}{60} to \nuc{Al}{26} flux ratio as this gauges the stellar evolution models and the age of the superbubbles.
    A flux ratio measured for the Local Bubble could further constrain models of \nuc{Fe}{60} deposition on Earth and its moon.
    }


   \maketitle
%

\section{Introduction}\label{sec:intro}
Measurements of the $\gamma$-ray line at 1809\,keV from decaying \nuc{Al}{26} suggest a quasi-persistent mass of this radioactive isotope of $1.2$--$2.4\,\mathrm{M_\odot}$ distributed throughout the Galaxy \citep{Pleintinger2023_26Al,Siegert2023_PSYCO}.
While it is commonly assumed that large parts of this mass is originating in massive stars \citep{Knoedlseder1999_freefree1.8MeV,Diehl2006_26Al}, also contributions from classical novae and AGB stars may play a role \citep[e.g.,][suggesting a nova contribution of up to 30\,\%]{Vasini2022_GCE}.
Being agnostic about the possible contribution from low-mass stars, \citet{Siegert2023_PSYCO} found a core-collapse supernova rate of $1.8$--$2.8$ per century based on comparisons with $\gamma$-ray measurements of the \nuc{Al}{26} decay line at 1809\,keV alone.
With their population synthesis model, also the mass of radioactive \nuc{Fe}{60} could be estimated, ranging between $1$--$6\,\mathrm{M_\odot}$.
Given the $\gamma$-ray line measurements of \citet{Wang2020_Fe60} who detected both decay lines of \nuc{Fe}{60} at 1173 and 1332\,keV in the Milky Way, this model estimate appears reasonable, but includes large uncertainties.

Data analyses of soft $\gamma$-ray emission, such as for nuclear lines, depend on spatial templates that are assumed to represent the emission.
Refining these models is then an iterative approach that can also lead to substantial changes in the understanding of the emission with ever-increasing exposure from the large field-of-view $\gamma$-ray telescopes.
In a recent work, trying to compare the raw data of the 1809\,keV line from INTEGRAL/SPI with Galactic-scale hydrodynamics simulations \citep[e.g.,][]{Fujimoto2018_26Alsim,RodgersLee2019_sim26Al,Krause2021_26Alchimneys}, \citet{Pleintinger2019_26Al} found a non-uniform scale height along Galactic longitudes.
In particular, the data suggest a scale height distribution which peaks around a few tens of pc and then stays rather flat up to, and possibly beyond, 2\,kpc.
On the one hand, this was surprising as, typically, emission templates with uniform scale heights worked particularly well in $\gamma$-ray data analyses \citep[e.g.,][]{Knoedlseder1999_26AlCOMPTEL,Diehl2006_26Al,Wang2009_26Al,Siegert2016_Orion26Al,Siegert2017_PhD511,Pleintinger2019_26Al,Wang2020_Fe60}.
On the other hand, this can be interpreted as superbubbles which open up towards higher latitudes allowing \nuc{Al}{26} to flow towards regions of lower densities \citep{Krause2021_26Alchimneys}.

Another interpretation of the findings by \citet{Pleintinger2019_26Al} is local foreground emission that would be hard to detect as an almost-isotropic component for current coded aperture mask $\gamma$-ray telescopes \citep{Siegert2022_gammaraytelescopes}.
However, the hints for very large scale heights may point to emission very nearby that is not directly connected to the Galactic background.
A natural candidate for such a foreground emission would be the Local Bubble \citep[e.g.,][]{Breitschwerdt1996_LocalBubble}, the superbubble in which the Solar System is currently located.

\begin{figure*}[!t]
    \centering
    \includegraphics[width=\columnwidth]{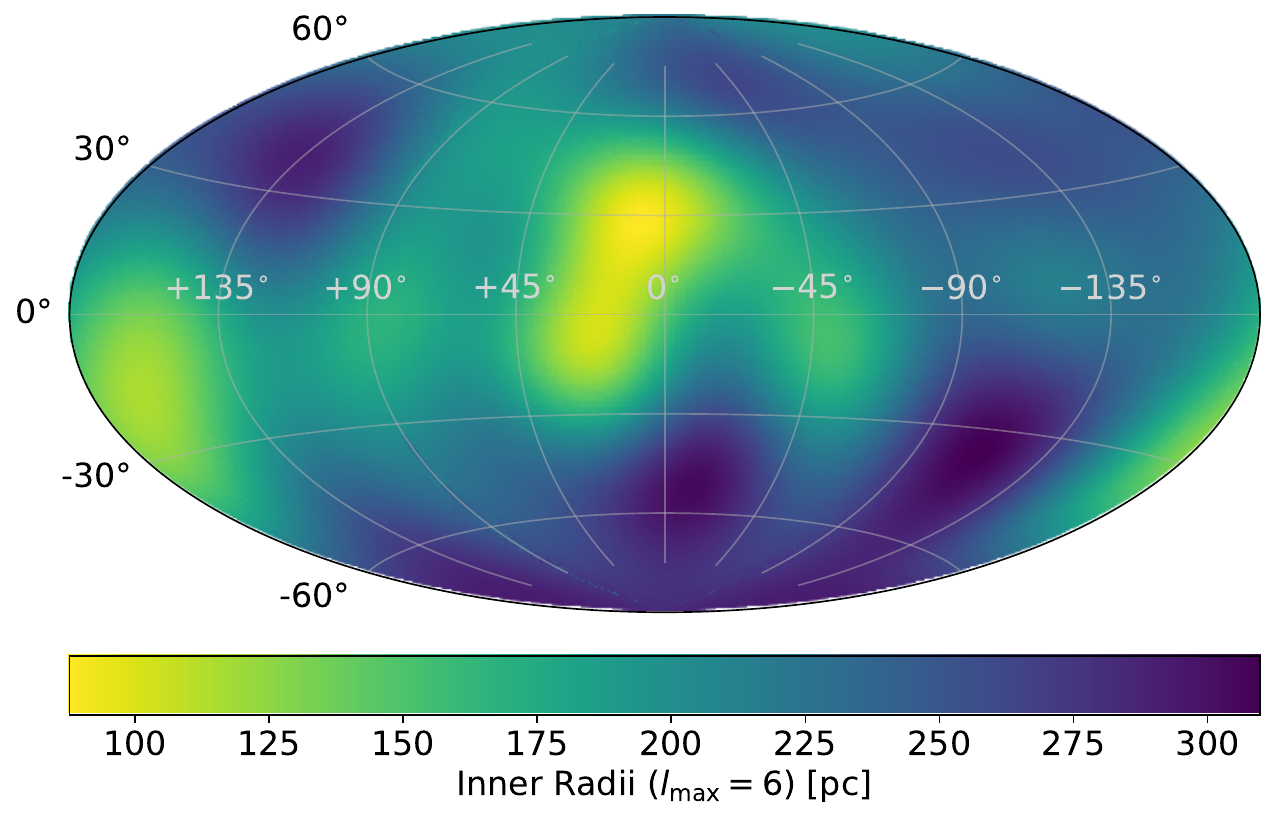}
    \includegraphics[width=\columnwidth]{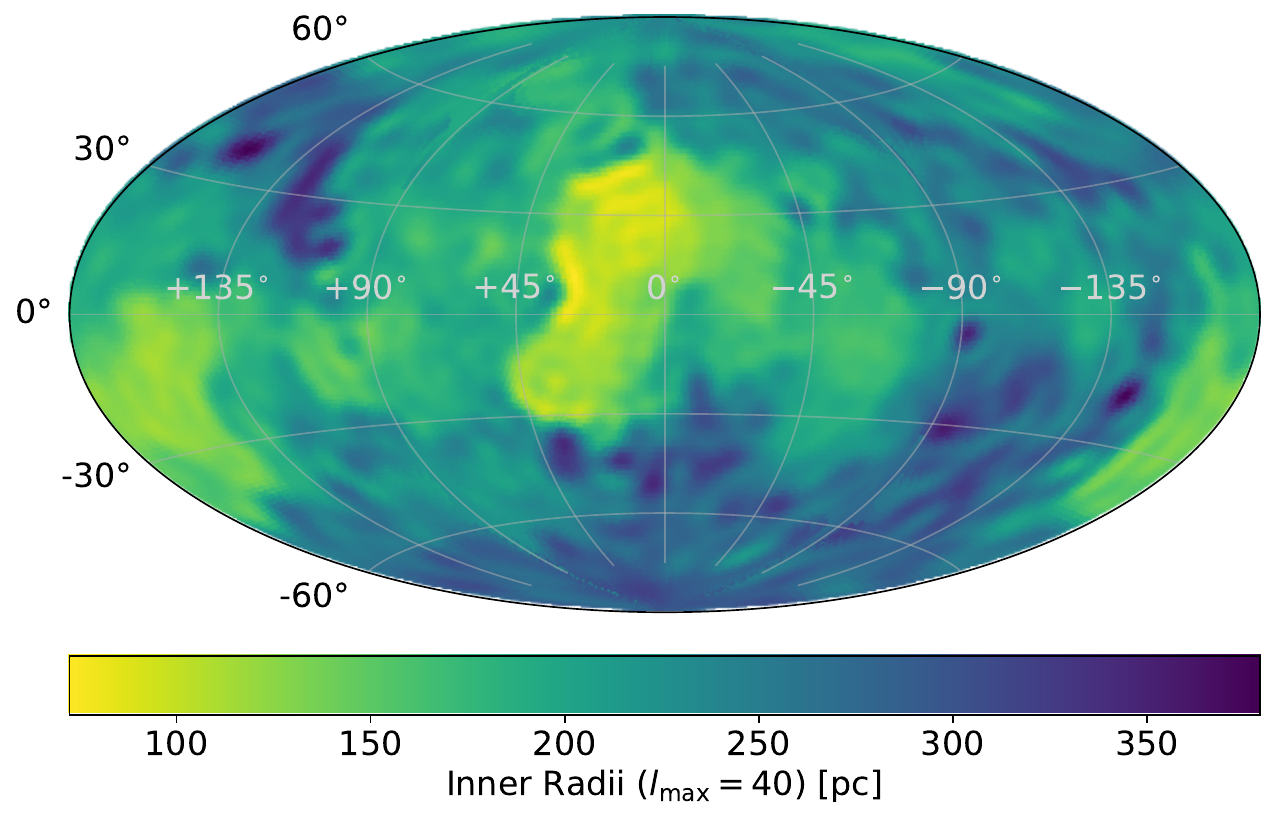}
    \caption{Inner boundaries of the geometrical Local Bubble model as derived by \citet{Pelgrims2020_LocalBubble}. Shown are two different granularities of the spherical harmomics decomposition with $l_{\rm max} = 6$ (left) and $l_{\rm max} = 40$ (right).}
    \label{fig:inner_radii}
\end{figure*}

This argument is further strengthened by measurements in deep-sea archives that show exceptionally high concentrations of radioactive \nuc{Fe}{60} particles in certain layers that cannot have been produced on Earth \citep[e.g.,][]{Wallner2016_Fe60,Wallner2021_oceancrust}.
This points to nearby supernova activity within a period of a few Myr, similar to the decay times of \nuc{Fe}{60} and \nuc{Al}{26}.
If there had been supernovae within the Local Bubble not too long ago, residual \nuc{Fe}{60} and \nuc{Al}{26} would still be present and currently decaying, leading to $\gamma$-ray emission from all directions.
In addition, due to the $\beta^+$-decay of \nuc{Al}{26}, positrons would currently be produced and would presumably annihilate in the bubble walls, creating another $\gamma$-ray line at 511\,keV.

In this study, we want to investigate how strong the $\gamma$-ray line emission from the Local Bubble is at the decay energies of 1809\,keV (\nuc{Al}{26}), 1332 and 1173\,keV (\nuc{Fe}{60}), and 511\,keV (positron annihilation).
In particular, we study the emission of $\gamma$-rays owing to the following decay chains and reactions, in which all the lifetimes ($\tau$) and probabilities ($p$) relevant for this work are indicated:
\begin{align}
    ^{60}\mathrm{Fe} & \xrightarrow[\beta^-]{\tau_{60} = 3.8\,\mathrm{Myr}} & ^{60}\mathrm{Co}^* & + e^- + \bar{\nu}_e \nonumber\\
    ^{60}\mathrm{Co}^* & \xrightarrow[\mathrm{IT}]{\tau = 10\,\mathrm{min}} & ^{60}\mathrm{Co} & + \gamma(59\,\mathrm{keV}\,;\,p=0.0207) \nonumber\\
    ^{60}\mathrm{Co} & \xrightarrow[\beta^-]{\tau = 7.6\,\mathrm{yr}} & ^{60}\mathrm{Ni}^* & + e^- + \bar{\nu}_e \nonumber\\
    ^{60}\mathrm{Ni}^* & \xrightarrow[\mathrm{IT}]{\tau \sim \mathrm{ps}} & ^{60}\mathrm{Ni} & + \gamma(1173\,\mathrm{keV}\,;\,p=0.9985)\nonumber\\
    & & & + \gamma(1332\,\mathrm{keV}\,;\,p_{60}=0.9998)\label{eq:fe60-decay}\\
    \hline\nonumber
    ^{26}\mathrm{Al} & \xrightarrow[\beta^+]{\tau_{26} = 1.03\,\mathrm{Myr}} & ^{26}\mathrm{Mg}^* & + e^+(p_{\beta+}=0.8201) + \nu_e \nonumber\\
    ^{26}\mathrm{Mg}^* & \xrightarrow[\mathrm{IT}]{\tau = 0.48\,\mathrm{ps}} & ^{26}\mathrm{Mg} & + \gamma(1809\,\mathrm{keV}\,;\,p_{26}=0.9976)\label{eq:al26-decay}\\
    \hline\nonumber
    e^+ + ~^1\mathrm{H} & \xrightarrow[\mathrm{Ch.~ex.}]{\tau \lesssim 1\,\mathrm{Myr}} &  \mathrm{Ps} & + p \nonumber\\
    \mathrm{para{-}Ps} & \xrightarrow[\mathrm{decay}]{\tau = 0.125\,\mathrm{ns}} & & 2\gamma(511\,\mathrm{keV}) \label{eq:pPs-decay}\\
    \mathrm{ortho{-}Ps} & \xrightarrow[\mathrm{decay}]{\tau = 142\,\mathrm{ns}} & & 3\gamma(\leq 511\,\mathrm{keV}) \label{eq:oPs-decay}\\
    \hline\nonumber
\end{align}
Because the probabilities of the final \nuc{Ni}{60} de-excitation in Eq.\,(\ref{eq:fe60-decay}) are very similar and close to $1.0$, we use the 1332\,keV line as surrogate for the expected \nuc{Fe}{60} emission.
We note that in data analyses, the significance of both lines combined is increased roughly as $\sqrt{2}$ (see Sect.\,\ref{sec:COSI_sims}).
The annihilation of positrons with electrons in the environment of the Local Bubble is assumed to be dominated by charge exchange (Ch.~ex.) with hydrogen, leading to the intermediate bound state of Positronium (Ps).
Depending on the spin state of Ps, either para-Ps or ortho-Ps decays on the nano-second timescale which results in two 511\,keV photons for only para-Ps, and a continuous spectrum up to 511\,keV for ortho-Ps \citep{Ore1949_511}.
The direct annihilation with (free) electrons is subdominant, and the cross section for radiative recombination with electrons inside the Local Bubble is several orders of magnitude smaller than for charge exchange.
For this reason, we only focus on the most dominant photon emission process in the case of positron annihilation, Eq.\,(\ref{eq:pPs-decay}), and discuss a correction factor for direct annihilation in Sects.\,\ref{sec:case1_positrons} and \ref{sec:case2_positrons}.

We base our estimates on two modelling assumptions, one from a geometrical point of view \citep{Pelgrims2020_LocalBubble,Zucker2022_LocalBubble} with physical arguments, and one from a hydrodynamics point of view \citep{Schulreich2023_LB} with detailed setups, both tailored to match the terrestrial deposition of \nuc{Fe}{60}.
By identifying the total fluxes, their isotropic contribution, and flux ratios as a function of position, we can suggest figures of merit of how to distinguish between different models, and also if it is indeed possible to perform these measurements with future $\gamma$-ray telescopes.

This paper is structured as follows:
In Sect.\,\ref{sec:LocalBubble}, we describe previous measurements of the Local Bubble considering size, shape, age, and radioactivity content, as well as recent hydrodynamics simulations tailored to match the radioactive impact on Earth.
Sect.\,\ref{sec:gammaray_expectations} includes the expected photon emission from different $\gamma$-ray lines originating from the Local Bubble, including \nuc{Al}{26}, \nuc{Fe}{60}, and the 511\,keV line considering two scenarios.
We describe the signal-to-Galactic-background ratio in Sect.\,\ref{sec:results} to determine emission hot spots and isotropic components.
Since current instruments are nearly-incapable of measuring isotropic emission, we estimate the detectability of the Local Bubble in these $\gamma$-ray lines with the future COSI-SMEX satellite instrument and present the results in Sect.\,\ref{sec:COSI_sims}.
Finally, we discuss our findings in Sect.\,\ref{sec:discussion} and conclude in Sect.\ref{sec:conclusion}.

\begin{figure*}[!ht]
    \centering
    \includegraphics[width=0.98\columnwidth]{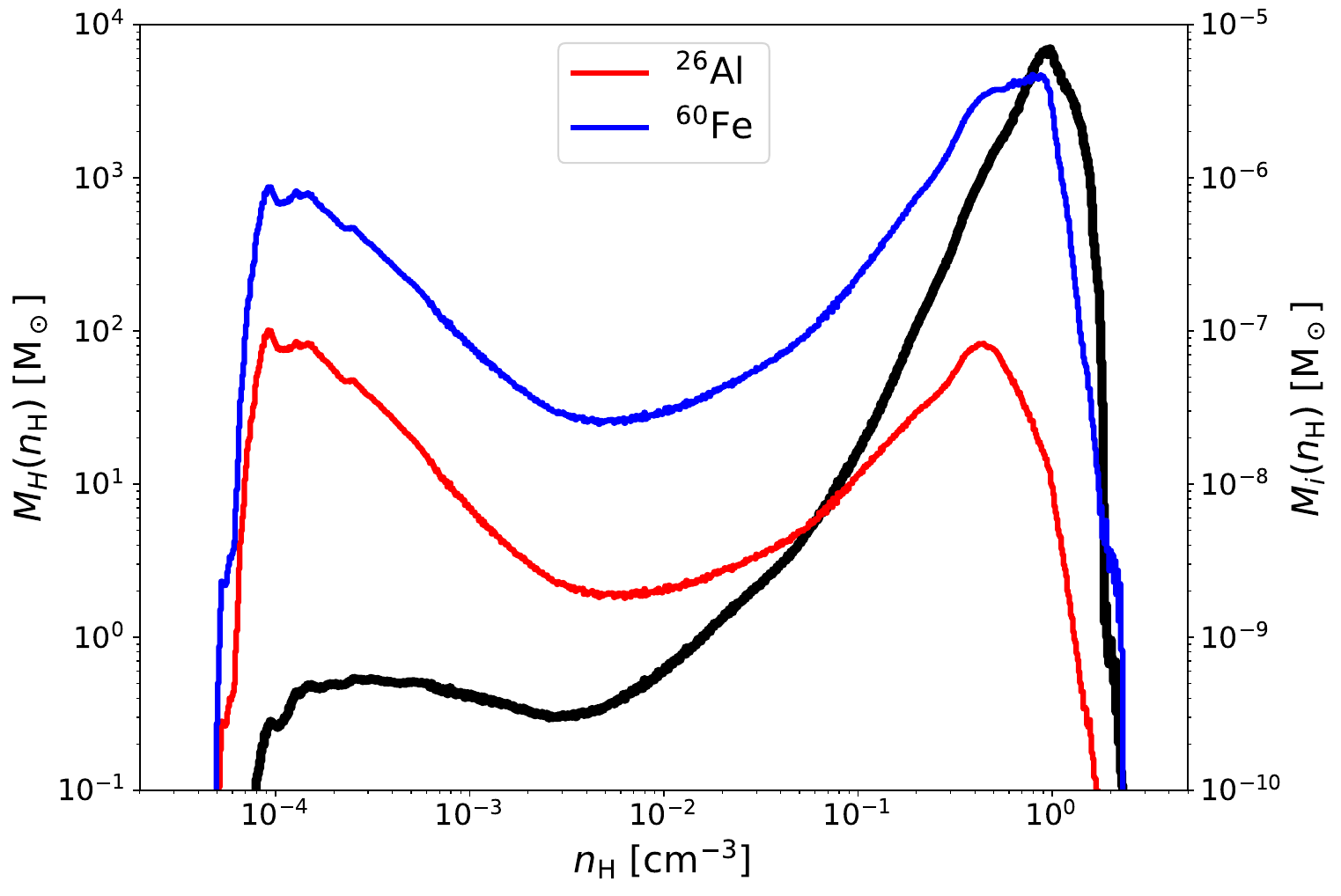}~
    \includegraphics[width=0.87\columnwidth]{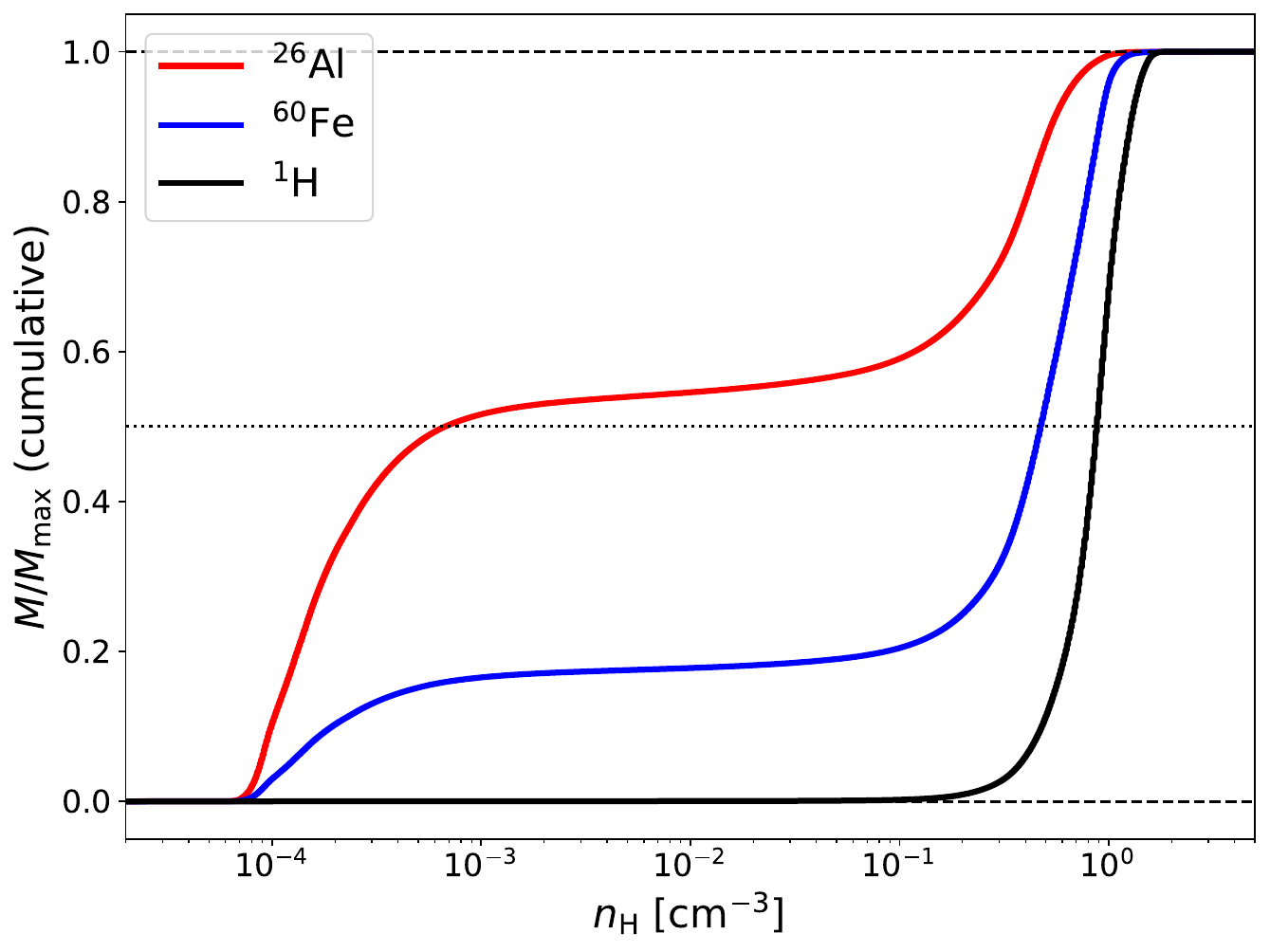}
    \caption{Left: Mass distributions of hydrogen (black; left axis), and \nuc{Al}{26} (red) and \nuc{Fe}{60} (blue; both isotopes right axis) as a function of $n_{\rm H}$ in the Local Bubble hydrodynamics simulation by \citet{Schulreich2023_LB}. Note that the right axis is scaled by $10^{-9}$ to the left axis to allow a visual comparison. Right: Cumulative distribution of the hydrogen, \nuc{Al}{26} and \nuc{Fe}{60} masses from the left panel with maximum (total) masses of $3.5 \times 10^{5}\,\mathrm{M_\odot}$, $1.6 \times 10^{-5}\,\mathrm{M_\odot}$ and $4.9 \times 10^{-4}\,\mathrm{M_\odot}$, respectively. Only cells for which the flow speed is higher than $1\,\mathrm{km\,s^{-1}}$ are used to ensure that the static background medium is not considered as part of the Local Bubble.
    }
    \label{fig:mass_distribution_per_nH}
\end{figure*}

\section{Local Bubble and Solar vicinity}\label{sec:LocalBubble}
The Local Bubble is an asymmetric superbubble, that is, a cavity of hot plasma surrounded by a shell of hydrogen and dust, in which the Solar System is currently propagating through.
It has been formed by massive stars and supernovae within the last $\sim 20$\,Myr.
Determining the three-dimensional (3D) geometry of the Local Bubble is not trivial \citep[e.g.,][]{Lallement2014_OB}, as it relies on accurate measurements of lines of sight to stars in front of and behind the Local Bubble walls.
As the number of stars towards higher latitudes becomes smaller, the estimates of the Local Bubble size in different directions becomes more uncertain.
In this paper, we use two assumptions for the geometry of the Local Bubble: (1) A measurement-based geometrical model that determines the inner radii of the Local Bubble using a spherical harmonics decomposition \citep{Pelgrims2020_LocalBubble}, and (2) hydrodynamics simulations in which the Local Bubble self-consistently forms in an inhomogeneous background medium through feedback processes of those massive stars that perished in nearby stellar populations calculated back over the last 20\,Myr \citep{Schulreich2023_LB}.
In the following, we briefly describe these two models.

\subsection{Recent measurements}\label{sec:LB_measurements}
\citet{Pelgrims2020_LocalBubble} used 3D dust density maps from \citet{Lallement2019_L19} that covers a volume which contains the entire Local Bubble to estimate the inner and outer boundaries of the bubble.
In their work, they base the analysis and modelling on the constructed 3D map of dust reddening from Gaia DR2 photometric data in combination with 2MASS measurements to estimate the dust extinction towards stars in all directions.
This results in a map that includes the differential extinction as a function of the distance to the Sun, which in turn can be used to estimate the gas density of the Local Bubble.
With a narrow sampling of lines of sight from the Sun outwards, \citet{Pelgrims2020_LocalBubble} could identify the inner and outer radii of the Local Bubble walls for all viewing angles.
The authors note that the outer radii are, sometimes, not reliable, for which reason we only use the inner radii in the following sections.
Depending on the direction, the Local Bubble shell thickness ranges between 50 and 150\,pc.
For inner radii between 80 and 360\,pc (Fig.\,\ref{fig:inner_radii}), this may be considered rather large as typical wall thicknesses range around 10--30\,\% \citep[e.g.,][]{Krause2013_superbubbles,Krause2014_superbubbles}.
Interestingly, \citet{Pelgrims2020_LocalBubble} found that the cavity appears to be closed in all directions, rather than showing a chimney-like structure which had been suggested in different studies about superbubbles in general \citep[e.g.,][]{Krause2021_26Alchimneys}.
In a next step, the measurements of the inner radii are expanded into spherical harmonics with different maximum multipole degrees $l_{\rm max}$ that adjust the level of complexity of the inner Local Bubble surface.
For a given $l_{\rm max}$, the internal structure of the Local Bubble are more or less pronounced, and the authors suggest $l_{\rm max} = 6$ for further analyses.
We will use two cases, 1a with $l_{\rm max} = 40$ and 1b with $l_{\rm max} = 6$, to illustrate the differences in these assumptions in Sect.\,\ref{sec:case1_general}.

Based on the work by \citet{Pelgrims2020_LocalBubble}, \citet{Zucker2022_LocalBubble} discuss the shapes and motions of dense gas and young stars inside and in the vicinity of the Local Bubble.
They find that, apparently, all young star-forming regions are close to the surface of the Local Bubble, and that these show an outward motion.
To explain this expansive movement, \citet{Zucker2022_LocalBubble} suggest a star-burst event near to what is now the centre of the Local Bubble about 14\,Myr ago.
The subsequent supernovae formed the Local Bubble that is now fragmenting near its boundaries to molecular clouds, hosting the next generations of young stellar clusters.
Such a picture is widely known as triggered star formation by stellar winds and supernovae.

\begin{figure*}[!ht]
    \centering
    \includegraphics[width=\textwidth]{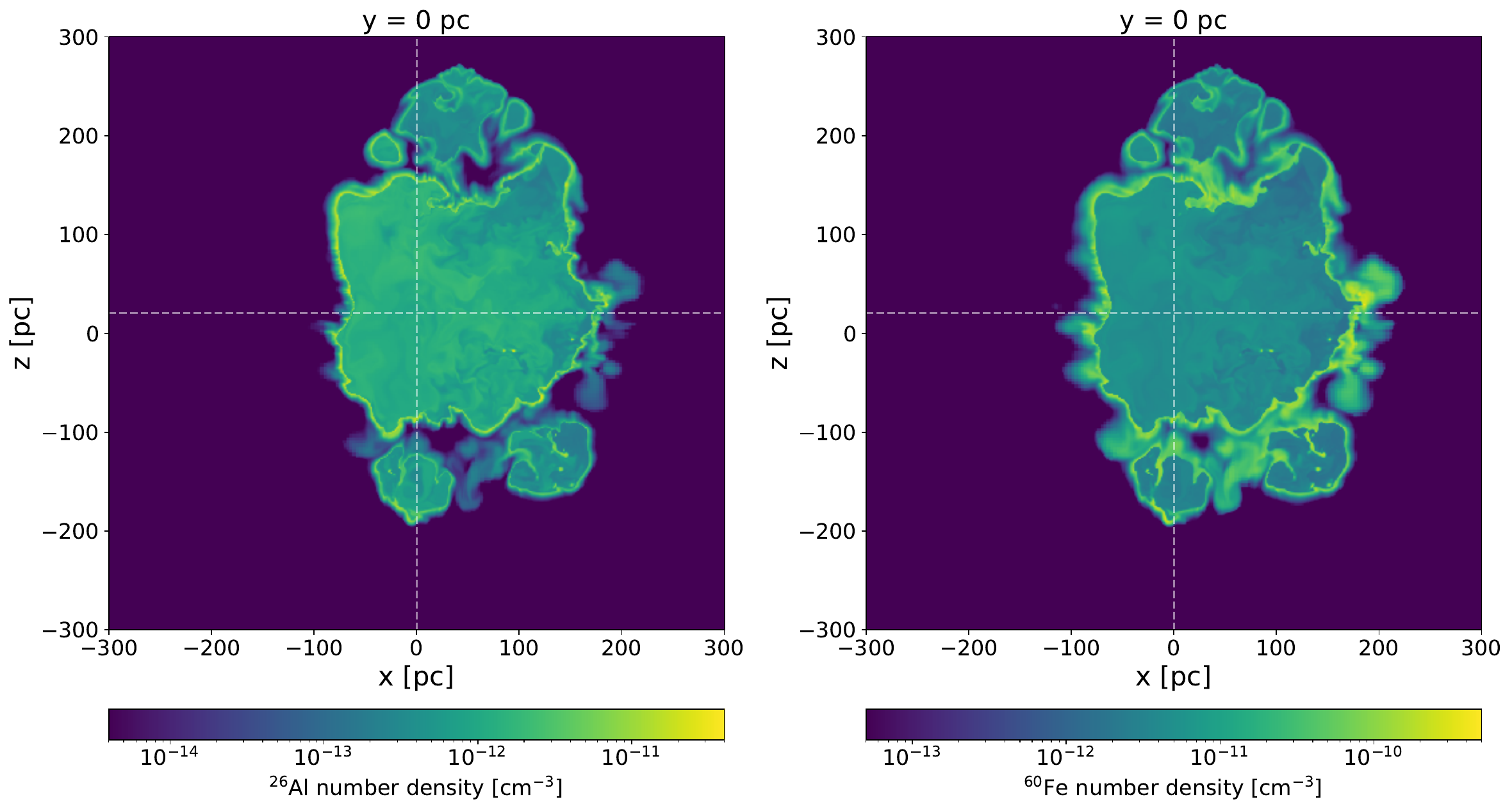}
    \caption{Slices at $y=0$\,pc through the \citet{Schulreich2023_LB} simulation for the number densities of \nuc{Al}{26} (left) and \nuc{Fe}{60} (right). It is evident that \nuc{Fe}{60} is mostly found along the bubble walls whereas \nuc{Al}{26} has also a large contribution of mass in the bubble's hot phase. The position of the Sun is marked with dashed gray lines at $z=20.8$\,pc.}
    \label{fig:MS_slice_26Al_60Fe}
\end{figure*}

\subsection{Recent hydrodynamics simulations}\label{secLLB_simulations}
The high-resolution 3D hydrodynamics simulations of \cite{Schulreich2023_LB} utilised in this work are an extensive update of those first presented in \cite{2015PhDT.......692S} and are based on the most sophisticated initial conditions for the formation and evolution of the Local Bubble determined to date.
These include the number of core-collapse supernova explosions, as well as when and where they occurred, relying on a Gaia EDR3-based stellar census of the Scorpius-Centaurus (Sco-Cen) OB association by \cite{2022AJ....163...24L}, which has already emerged as the most likely source of most, if not all, of these supernovae in previous studies \citep[e.g.][]{2006MNRAS.373..993F}.
By fitting the initial mass function (IMF) of \cite{2001MNRAS.322..231K}, the total number of missing and thus exploded stars was found to be 14, of which 13 occurred in Upper Centaurus-Lupus and Lower Centaurus-Crux (UCL/LCC), and one in V1062~Sco (both being Sco-Cen populations).
The lifetimes of the perished stars and therefore the timing of their explosions -- assuming that all members of a population are formed simultaneously -- were estimated by interpolating between the rotating stellar evolution tracks for solar metallicity of \cite{2012A&A...537A.146E} to match the stars' initial masses, obtained from IMF binning.
For deriving the trajectories of the supernova progenitors, a novel Monte Carlo-type approach was developed, for which 10\,000 realisations of each progenitor population were back-calculated in time by means of test-particle simulations with a realistic Milky Way potential \citep{2016A&A...593A.108B}.
The explosion sites were selected based on the maximum values in six-dimensional phase-space probability distributions constructed from the stellar tracebacks.

The hydrodynamics simulations studied the turbulent transport of the radioisotopes \nuc{Al}{26}, \nuc{Mn}{53}, \nuc{Fe}{60}, and \nuc{Pu}{244}\footnote{While no $\gamma$-ray line is associated with the decay of \nuc{Mn}{53}, there are two potentially detectable ones at 554.6 and 597.4 keV for \nuc{Pu}{244}, stemming from the \nuc{Np}{240} decay to \nuc{Pu}{240}. However, since we currently always consider the Local Bubble in isolation, i.e.~without all the other superbubbles flanking it, which could also have swept up pre-existing \nuc{Pu}{244}, we do not consider it worthwhile to carry out \nuc{Pu}{244} line-of-sight integrations in the context of this paper.} due to stellar wind and supernova activity in the domain of the present-day Local Bubble, which forms over the last 20\,Myr (the birth time of the two progenitor populations) in a smoothly stratified modelled local interstellar medium.
With all the radioisotopes treated as decaying passive tracers, only \nuc{Al}{26} is released by the stellar winds (set up to be age- and initial mass-dependent), for which the yields of \cite{2012A&A...537A.146E} were used.
The supernova yields for \nuc{Al}{26}, \nuc{Mn}{53}, and \nuc{Fe}{60}, on the other hand, were taken from \citet{Limongi2018_LC18}.
Only \nuc{Pu}{244} was assumed to be pre-seeded (possibly by a kilonova event prior to the formation of the Local Bubble) in the sense of a two-step scenario \citep[see also e.g.][]{2021ApJ..923...219W}, with its initial concentration reconstructed from fitting the measurements of \citet{Wallner2021_oceancrust} in the uppermost deep-sea crust layers.
The Solar System, which, like each of the 14 supernova progenitors, travels as a `stellar particle' along its pre-calculated trajectory, crosses the outer shell of the Local Bubble about 4.6\,Myr ago and acts as a `moving detector' for the radioisotopic fluxes originating from the stellar feedback processes throughout the entire simulation time.
Without any real fine-tuning (only the density in the Galactic mid-plane was varied and finally set to 0.7\,H\,cm$^{-3}$), the numerical calculations are able to reproduce remarkably well the measurements of the four radioisotopes currently available for the time period from the Sun's entry into the Local Bubble.
In addition, its present-day extent and the value of the thermal pressure of the hot plasma in its interior, which was estimated by \cite{2014ApJ...791L..14S} from a combination of disparate observational results, are matched.
An even better match to the $\sim$3-Myr-old \nuc{Fe}{60} signal was found when the massive star responsible for the most recent supernova (about 0.88\,Myr ago) was removed from the simulation, implying a slight deviation from the most probable initial mass spectrum underlying the original scenario.
We consider the 14-supernovae scenario as case 2a and the alternative simulation in which the last supernova was simply taken out as case 2b.

\begin{figure*}[!ht]
    \centering
    \includegraphics[width=\textwidth]{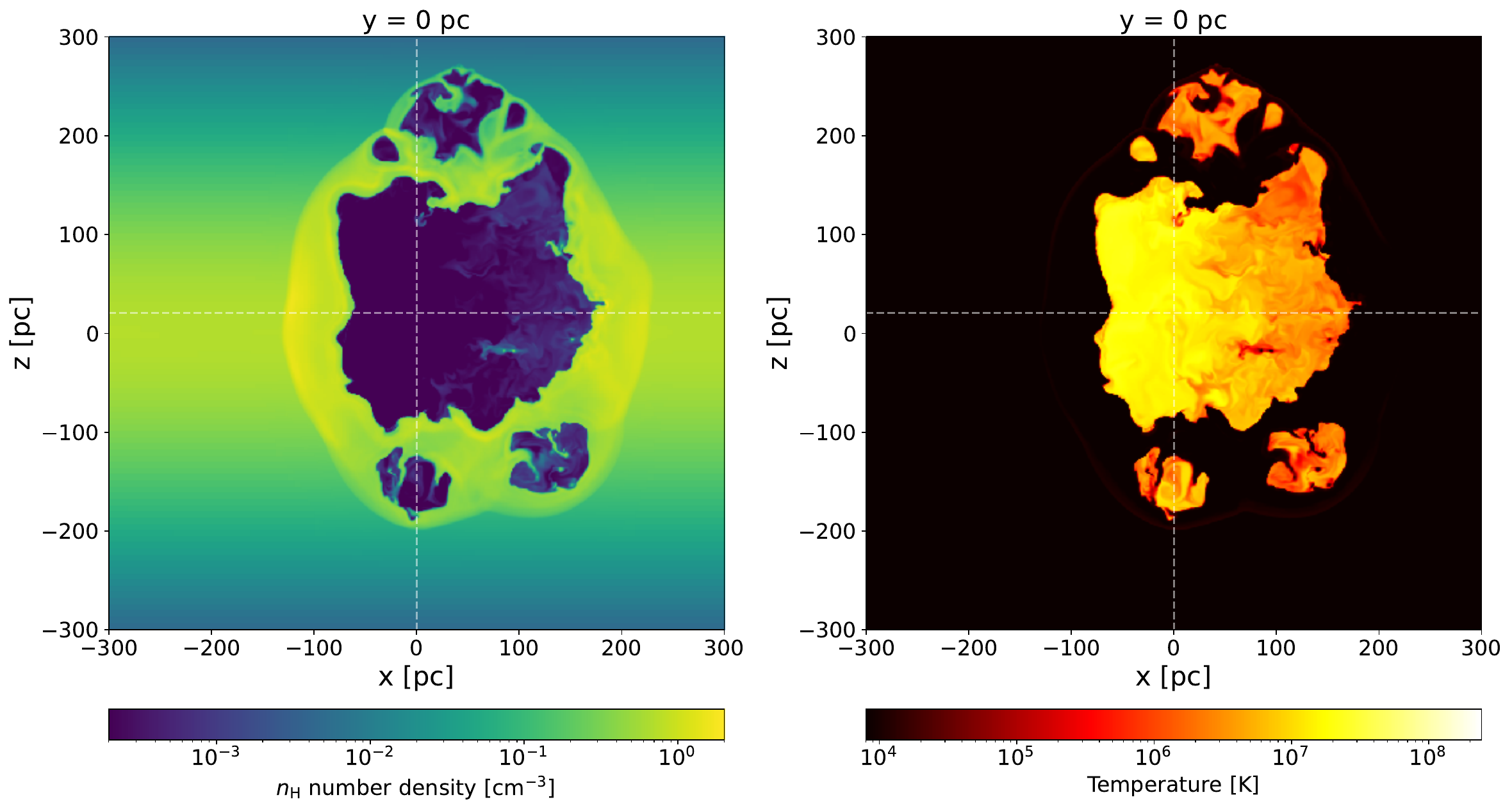}
    \caption{Slices at $y=0$\,pc through the \citet{Schulreich2023_LB} simulation for the neutral hydrogen number density (left) and the temperature (right). The gas outside the Local Bubble is shaped according to the Galactic plane density distribution, decreasing exponentially towards higher $|z|$. Small clumps of higher-density gas inside the bubble are visible in both panels, for example around $(x,z) = (80,-20)$\,pc. The hot gas is naturally correlated with thin phases of the ISM, and the bubble walls are at a temperature of $8000$\,K or below. The position of the Sun is marked with dashed gray lines at $z=20.8$\,pc.}
    \label{fig:MS_slice_nH_Temp}
\end{figure*}

\section{$\gamma$-ray line emission from the Local Bubble}\label{sec:gammaray_expectations}
For the \nuc{Al}{26} and subsequent 511\,keV emission, and \nuc{Fe}{60} $\gamma$-ray line emission, we are considering two case studies:
In the first case (case 1), we use the geometric model as inferred from measurements by \citet{Pelgrims2020_LocalBubble} \citep[see also][]{Zucker2022_LocalBubble}, in combination with the estimated \nuc{Fe}{60} deposit as determined in \citet{Chaikin2022_60Fe_LocalBubble}, to calculate the emissivity of \nuc{Al}{26}, its decay positrons, and \nuc{Fe}{60} $\gamma$-rays.
The second case (case 2) is based on hydrodynamics simulations by \citet{Schulreich2023_LB}, who tailored their simulations to match not only the \nuc{Fe}{60} deposit on Earth but in addition \nuc{Al}{26}, \nuc{Mn}{53}, and \nuc{Pu}{244} influxes from past supernovae in the Solar neighbourhood.
The premises for both cases are vastly different, so that the final emissivities and $\gamma$-ray flux maps will provide a reasonable range of possibilities of how the Local Bubble may look like as seen from an observer inside at the position of the Sun.
We will discuss the two cases separately in the following, for the three $\gamma$-ray lines considered, 1809\,keV from \nuc{Al}{26}, 1332\,keV from \nuc{Fe}{60} (the flux and appearance from the 1173\,keV line from \nuc{Fe}{60} is identical), and 511\,keV from Ps-decay.

\subsection{Case 1: Geometric model and \nuc{Fe}{60} deposit}\label{sec:case1_general}
The geometric model of \citet{Pelgrims2020_LocalBubble} delivers the inner boundaries of the Local Bubble, described as a function of distance to the Sun in the centre of the coordinate system.
Their model is provided as a HEALPix array with a side length of 128, resulting in $196\,608$ values for the three Cartesian directions, $x$, $y$, and $z$.
\citet{Pelgrims2020_LocalBubble} provide different levels of granularity in the reconstruction of the inner surface of the bubble as described by an expansion of the measurements into spherical harmonics up to multipoles $l_{\rm max}$ of 2, 4, 6, 8, 10, 20, 30, and 40.
For illustration purpose in this work, we use $l_{\rm max} = 6$ and $40$.
The former is a reasonable trade-off between the granularity of the shell and the expected turbulence.
We use the latter as an extreme case, that should however not be over-interpreted because the maps of distances, $R(\ell,b)$, show some `ringing effect'\footnote{This effect is similar to `over-fitting', which results in arbitrarily good fits but may lack a sound interpretation.} (Pelgrims, priv. comm.), but which results in $\gamma$-ray flux maps that are more in line with the hydrodynamics simulations naturally including turbulence on the pc scale (see Sect.\,\ref{sec:case2_general}).

In what follows, we will always construct all-sky maps on a $1^\circ \times 1^\circ$ (rectangular) pixel grid, that is, we will calculate the differential flux in units of $\mathrm{ph\,cm^{-2}\,s^{-1}\,sr^{-1}}$ for each of the $360 \times 180 = 64\,800$ pixels and each process.
In case 1 this means we need to define an emissivity $\epsilon(x,y,z)$ in units of $\mathrm{cm^{-3}\,s^{-1}}$ inside the inner boundaries for \nuc{Fe}{60} and \nuc{Al}{26}, and the outer boundaries for 511\,keV.
The first step is then interpolating the maps of radii from HEALPix in Cartesian into spherical coordinates spanned by the line of sight variable $s$ (from the point of the observer to infinity) and the two Galactic coordinates of longitude $\ell$ and latitude $b$.
The emissivity profiles that we define (see below) are then also transformed from Cartesian to spherical coordinates, $\epsilon(x,y,z) \rightarrow \epsilon(s,\ell,b)$, from which the line-of-sight integration is performed as
\begin{equation}
    F(\ell,b) = \frac{1}{4\,\pi\,\mrm{sr}} \int_0^\infty\,\mrm{d}s\,\epsilon\left(\Tilde{x}(s,\ell,b),\Tilde{y}(s,\ell,b),\Tilde{z}(s,\ell,b)\right)\mrm{,}
    \label{eq:los_case1}
\end{equation}
where the line of sight is defined as
\begin{align}
    \Tilde{x}(s,\ell,b) & =  x_\odot + s\,\cos \ell\,\cos b\,, \nonumber\\
    \Tilde{y}(s,\ell,b) & =  y_\odot + s\,\sin \ell\,\cos b\,, \nonumber\\
    \Tilde{z}(s,\ell,b) & =  z_\odot + s\,\sin b\,,
    \label{eq:los_coordinates}
\end{align}
with $\left(x,y,z\right)_\odot$ being the coordinates of the Sun which is taken here to be the position of the observer.
The coordinates are chosen so that the positive $x$-direction points towards the Galactic centre, the positive $z$-direction to the Galactic north pole, and the positive $y$-direction to $\ell = 90^\circ$.
\citet{Pelgrims2020_LocalBubble} used $\left(x,y,z\right)_\odot = \left(0,0,0\right)$ as the coordinates of the Sun, which we keep for case 1a and 1b \citep[for a more realistic Sun position, see][see also Sect.\,\ref{sec:case2_general}]{Bennett2018_Sun_GaiaDR2}.
The comparison between the geometrical model and the hydrodynamics simulation is then only biased by the observer position with a vertical difference of $20.8$\,pc.

The two geometries of $l_{\rm max} = 40$ and $l_{\rm max} = 6$ differ only slightly, with their minimum and maximum radii being larger and smaller, respectively.
We refer to cases 1a and 1b for $l_{\rm max} = 40$ and $l_{\rm max} = 6$ in the following but will restrict the illustrations to case 1a.

\subsubsection{\nuc{Fe}{60} and \nuc{Al}{26} emissivities}\label{sec:case1_nuc_emissivities}
We calculate the luminosity of a radioactive isotope $i$ inside the Local Bubble as follows.
Given its total ejecta mass $M_i$, its lifetime $\tau_i$, its atomic mass $m_i$, and the probability to emit a $\gamma$-ray photon $p_i$, we find the total luminosity $L_i(t)$ as a function of time $t$ as
\begin{equation}
    L_i(t) = \frac{M_i\,p_i}{m_i\,\tau_i}\,\exp\left(-\frac{t-t_0}{\tau_i}\right) \equiv L_{0,i}\,\exp\left(-\frac{t-t_0}{\tau_i}\right)\,,
    \label{eq:luminosity_nuc}
\end{equation}
where $t_0$ is the time at which the isotope is ejected (or produced).
Since the Local Bubble is not a distant point source, we need to take into account both, the expected profile of isotope $i$ inside, and the boundaries of the bubble to identify the limits of the line-of-sight integration.
It has been suggested that the ejected mass from supernovae and stellar winds in superbubbles quickly homogenise on a timescale of $\sim$$\mathrm{Myr}$ \citep{Krause2013_superbubbles,Krause2014_superbubbles} from which a constant emissivity could be expected.
Here, we use a slightly more complex model and convert the time-dependence of Eq.\,(\ref{eq:luminosity_nuc}) into a radial dependence using a typical sound velocity inside the bubble of $\varv_{\rm turb} \sim 300\,\mathrm{km\,s^{-1}}$, from which we define the emissivity for the isotope $i$ as
\begin{equation}
    \epsilon_i(r) = \epsilon_{0,i}\,\exp\left(-\frac{t_{\rm SN}}{\tau_i}\right)\,\exp\left(-\frac{r(x_{\rm SN},y_{\rm SN},z_{\rm SN})}{\tau_i\,\varv_{\rm turb}}\right)\,,
    \label{eq:emissivity_profile}
\end{equation}
where $\left(x,y,z\right)_{\rm SN}$ is the position of a supernova with age $t_{\rm SN}$ so that $r(x_{\rm SN},y_{\rm SN},z_{\rm SN}) = \sqrt{(\Tilde{x}(s,\ell,b) - x_{\rm SN})^2 + (\Tilde{y}(s,\ell,b) - y_{\rm SN})^2 + (\Tilde{z}(s,\ell,b) - z_{\rm SN})^2}$.
The normalisation factor $\epsilon_{0,i}$ is obtained by calculating the `effective volume' $V$ of the Local Bubble, having its boundary at $R(\ell,b)$,
\begin{equation}
    V = \int\mrm{d}\Omega\,\int_0^{R(\ell,b)}\mrm{d}s\,s^2\,\frac{\epsilon_i(r(s,\ell,b))}{\epsilon_{0,i}}\,,
    \label{eq:effective_volume}
\end{equation}
so that finally
\begin{equation}
    \epsilon_{0,i} = \frac{M_i\,p_i}{m_i\,\tau_i V} \equiv \frac{L_{0,i}}{V}\,.
    \label{eq:emissivity_norm}
\end{equation}

With the sound speed of $\sim$$ 300\,\mathrm{km\,s^{-1}}$, the lifetimes of \nuc{Al}{26} and \nuc{Fe}{60} of $\tau_{26} = 1.05\,\mathrm{Myr}$ and $\tau_{60} = 3.8\,\mathrm{Myr}$, respectively, and the typical radial scale of the Local Bubble between 100 and 300\,pc, the second exponential term in Eq.\,(\ref{eq:emissivity_profile}) amounts to a decrease from the centre to the edges of 0.7--0.4 in the case of \nuc{Al}{26}, and 0.9--0.8 for \nuc{Fe}{60}.
This means, there will be a large fraction of the emissivity distributed isotropically, but still some `hotspot' left, indicating of where the last supernova happened.
It is also clear already from this consideration that the distributions of \nuc{Al}{26} and \nuc{Fe}{60} are not completely co-spatial, as typically assumed in $\gamma$-ray data analyses \citep[e.g.,][]{Wang2007_60Fe,Wang2020_Fe60,Siegert2023_PSYCO}.
As will be discussed in Sect.\,\ref{sec:case2_general}, it is in fact possible to identify the position of the last supernova by hotspots in the 1809\,keV flux map.

In the model of \citet{Chaikin2022_60Fe_LocalBubble} two supernovae 3 and 7\,Myr ago are held responsible for the \nuc{Fe}{60} measured by \citet{Wallner2021_oceancrust} in deep-sea samples.
The positions of these two supernovae are not provided by the authors.
We mimic the position of the 3\,Myr-old supernova from the positions derived in \citet{Schulreich2023_LB} (see also Sect.\,\ref{sec:case2_general}) that occurred at $(x,y,z) = (40.9, -65.9, 19.5)$\,pc and about 92.2\,pc away at the time of the explosion, and the 7\,Myr-old supernova at $(x,y,z) = (113.1, -15.6, 7.3)$\,pc, about 220.9\,pc away.
Thus, we calculate two emissivity profiles by using Eq.\,(\ref{eq:emissivity_profile}) and simply add them.
We note that the mixing inside the Local Bubble and in particular its shape is bound to change within the last 7\,Myr.
For the sake of simplicity in this case 1, we nevertheless keep it at this straight-forward model, also because the models for $\gamma$-ray observations typically use these simple assumptions (see also the discussion about these assumptions in Sect.\,\ref{sec:discussion}).
While \citet{Chaikin2022_60Fe_LocalBubble} discuss the effects of the uptake of dust particles on Earth which would lead to a range of plausible ejecta masses, we keep the \nuc{Fe}{60} yield per supernova for our geometrical model to their canonical value of $10^{-4}\,\mathrm{M_\odot}$.
Considering this \nuc{Fe}{60} ejecta mass, we need to consider the possible impacts of the progenitor stars that may alter the yields of \nuc{Al}{26}.
These include the rotational velocity, the metallicity, and to a lesser extent the binarity.
The rotational velocities of the progenitor stars are unknown, but given that they must have been O- or B-type stars to form the Local Bubble, we can estimate an expectation value and a range of possible rotational velocities from the catalogue of \citet{Glebocki2005_rotvel_OB}.
We find a range of $140 \pm 90\,\mathrm{km\,s^{-1}}$ for O- and $135 \pm 105\,\mathrm{km\,s^{-1}}$ for B-type stars, which we combine into a range of $30$--$240\,\mathrm{km\,s^{-1}}$.
This covers large parts of the calculations by \citet{Limongi2018_LC18} with a velocity grid of $0$, $150$, and $300\,\mathrm{km\,s^{-1}}$.
The metallicity of these young objects are even less constrained:
Given the metallicity gradient of the Milky Way by \citet{Cheng2012_MW_metallicity}, for example, one could estimate the average metallicity around the Solar circle to be about $-\left[\mathrm{Fe/H}\right] = 0.29$--$0.90$.
However, since the Sun already has solar metallicity by definition, any much younger object might have super-solar metallicity, which would be difficult to estimate.
\citet{Limongi2018_LC18} calculated a grid of metallicities ranging between $10^{-3}$, $10^{-2}$, $10^{-1}$, and $10^0$, so that any super-solar yield would require an extrapolation from the available ones.
As a conservative approach, we will use a metallicity of $\left[\mathrm{Fe/H}\right] = 0$ to estimate the masses of the progenitor stars.
Using all the above arguments, we find that the two progenitor stars that exploded 3 and 7\,Myr ago must have had initial masses of $13$--$25\,\mathrm{M_\odot}$, assuming the yield model by \citet{Limongi2018_LC18}.
The resulting yields for \nuc{Fe}{60} and \nuc{Al}{26} are then $1.0 \times 10^{-4}\,\mathrm{M_\odot}$ (by definition) and $(1.6$--$13.0) \times 10^{-5}\,\mathrm{M_\odot}$, respectively.

\begin{figure}[!ht]
    \centering
    \includegraphics[width=1.0\columnwidth]{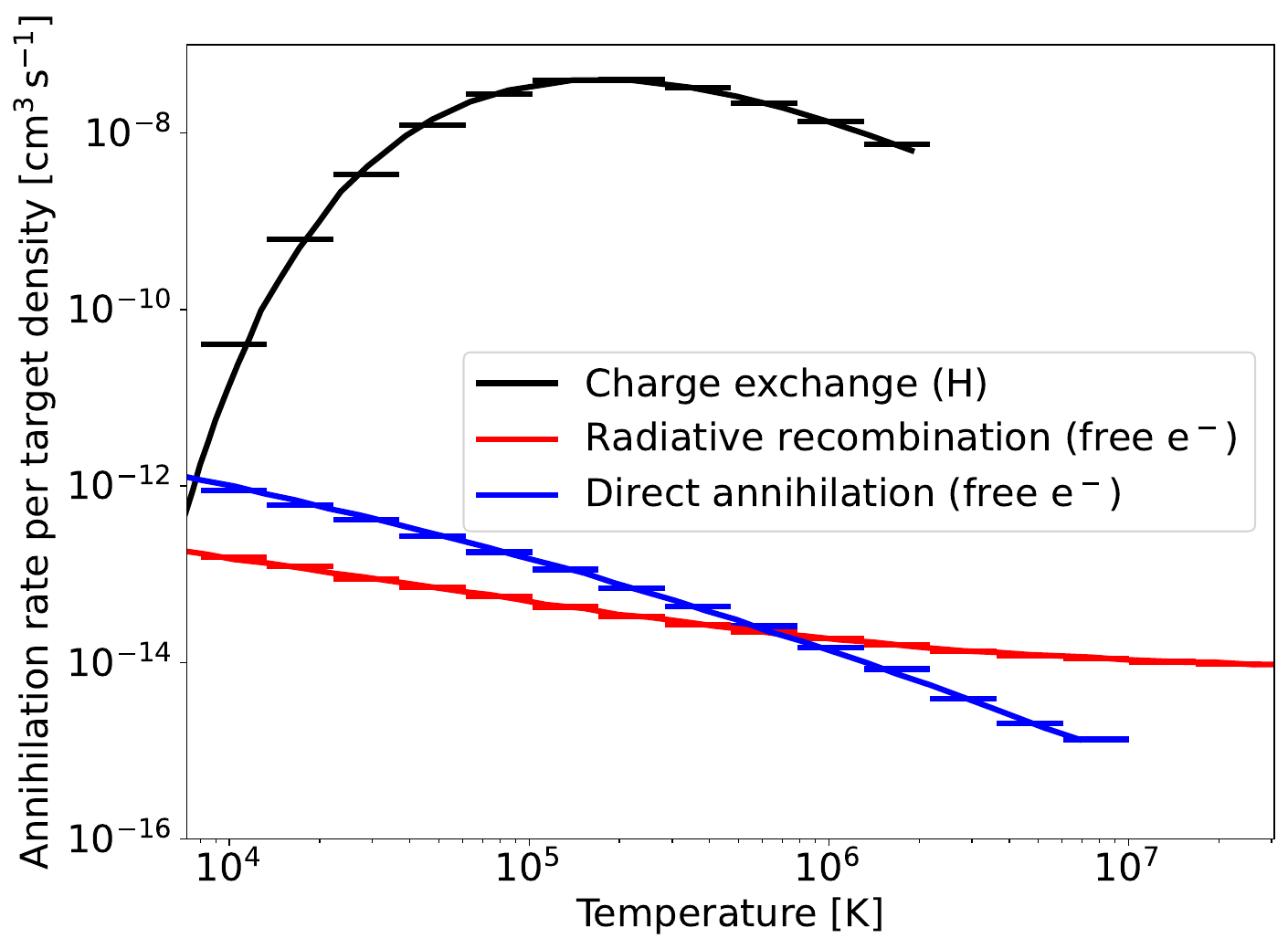}
    \caption{Positron annihilation rates per target density for the processes relevant in this study. The values (solid lines) are taken from \citet{Guessoum2005_511} and were rebinned (steps) for the purpose of assigning rates in the hydrodynamics simulation by \citet{Schulreich2023_LB}.}
    \label{fig:annihilation_rates}
\end{figure}
\subsubsection{Positron annihilation from \nuc{Al}{26}}\label{sec:case1_positrons}
The isotope \nuc{Al}{26} decays with a probability of $p_{\beta+} = 0.82$ via $\beta^+$-decay and thereby emits a positron with a mean kinetic energy of $543$\,keV in the $\beta$-decay spectrum up to an end point energy of $1173$\,keV.
The positron is therefore at most mildly relativistic with $\gamma \leq 3.3$.
Low-energy positron propagation is hardly understood, that is, if they propagate ballistically, diffusively, or if magneto-hydrodynamic waves have a large impact \citep[see][for details]{Jean2009_511ISM}.
From the energy losses, here mainly ionisation and Coulomb losses, as well as the annihilation rates at different temperatures of the interstellar medium \citep{Guessoum2005_511}, we estimate \emph{where} the positrons tend to annihilate in the context of a superbubble like the Local Bubble.

The path length of a positron from \nuc{Al}{26} with maximum Lorentz-factor of $\gamma_{\rm max} = 3.3$ inside a superbubble with a hydrogen density of $n_{\rm H,min} \sim 10^{-4}\,\mathrm{cm^{-3}}$ is on the order of $R_{\rm ball,max} \sim 100$\,Mpc -- inside of bubble walls with a density of $n_{\rm H,max} \sim 10^{1}\,\mathrm{cm^{-3}}$ on the order of $R_{\rm ball,min} \sim 1$\,kpc.
Positrons certainly propagate along the (possibly tangled) magnetic field lines in superbubbles, which reduces effective distances by magnetic diffusion.
We approximate the diffusion length scale by $R_{\rm diff} \approx \frac{D}{\kappa\, \varv_{+}}$, where $\kappa \sim 2$--$3$, depending on the directionality of the magnetic field to diffuse in, $D$ is the (unknown) diffusion coefficient, and $\varv_{+}$ is the Alfv\'{e}n speed.
We note that the Alfv\'{e}n speed differs in detail depending on the magnetic field strength, and the density; for an order of magnitude estimate, we will use the range of $10^1$--$10^3\,\mathrm{km\,s^{-1}}$.
Typical values of $D$ for GeV positrons range around $10^{27}\,\mathrm{cm^2\,s^{-1}}$, which we use here as canonically value also for mildly relativistic positrons \citep[see][for different scenarios on how the low-energy positron propagation might be realised in the interstellar medium]{Martin2012_511}, so that the diffusion length scale is on the order of $10^0$--$10^2$\,pc for this collisionless transport model.
The effective distance can then be approximated as $R_{\rm dist} \approx \sqrt{R_{\rm ball} \times R_{\rm diff}}$ which ranges from 1\,pc to several 100\,kpc.
We note that the propagation of low-energy positrons is not well understood and that different scenarios, such as collisionless transport, intermediate inhomogeneous transport, and pure ballistic transport may all be realised in nature \citep{Martin2012_511}.
Thus, in the latter case, positrons would diffuse through the Local Bubble walls into the next bubble, from which they might also escape, so that the annihilation emission may not be traced back to the production site.
In the first case, the positrons would annihilate rather quickly once they reach a density high enough to lose their kinetic energy efficiently.
As the last case may be interesting on a global, Galactic-wide, picture, we will only consider and discuss the first case in this work.

In order to estimate the emissivity profile of \emph{annihilating} positrons, we consider again the effect of diffusion, now in competition with the effect of annihilation.
The number density of \emph{annihilating} positrons can be described as a function of radial coordinate as
\begin{equation}
    \frac{1}{n^*(r)}\frac{\partial n^*(r)}{\partial t} = \dot{D}(r) + \dot{A}(r)\mathrm{,}
    \label{eq:diff_ann_equation}
\end{equation}
where $\dot{D}(r)$ is the diffusion term of a positron propagation from a spherical shell at $r$ into a shell at $r+dr$, and reads
\begin{equation}
    \dot{D}(r) = 4\,\pi\,\varv(r) \left[ n^*(r)\,r^2 - n^*(r+dr)\,(r+dr)^2\right]\,,
    \label{eq:diffusion_term}
\end{equation}
with $\varv(r) = \frac{D}{\kappa\,r}$ being the diffusion velocity.
In Eq.\,(\ref{eq:diff_ann_equation}), $\dot{A}(r)$ is the annihilation term which removes particles with an annihilation rate $\dot{a}(T)$ in a shell at $r$ with a hydrogen number density $n_{\rm H}(r)$, so that
\begin{equation}
    \dot{A}(r) = - 4\,\pi\,r^2\,\dot{a}(T(r))\,n_\mrm{H}(r)\,n^*(r)\,\mrm{d}r\,.
    \label{eq:annihilation_term}
\end{equation}
We note that the annihilation rates $\dot{a}$ \citep{Guessoum2005_511} per target density, that is, in units of $\mathrm{cm^3\,s^{-1}}$, will automatically depend on the position as well as the temperature changes across the bubble.
Assuming a steady state, i.e. setting the left hand side of Eq.\,(\ref{eq:diff_ann_equation}) to zero, we derive the general solution of the density of \emph{annihilating} positrons as
\begin{equation}
    n^*(r) = n_0^*\,\exp\left\{-2 \int_{r_0}^r\left[ \frac{1}{r} + \frac{\kappa\,\dot{a}(T(r))\,n_\mrm{H}(r)}{2\,D}\,r \right]\,\mrm{d}r \right\}\,.
    \label{eq:general_solution_deq}
\end{equation}
A short derivation of this general solution is found in Appendix \ref{app:diff_eq_solution}.
The impact of the diffusion coefficient is now estimated by solving the integral for different conditions of a superbubble.
The annihilation rates per target density, $\dot{a}(T(r))$, are chosen from \citet{Guessoum2005_511}; we restrict ourselves to the cases of charge exchange (Ch.~ex.) with hydrogen, radiative recombination (rad.~rec.) with electrons, and direct annihilation in flight (daf).
At a specific position with high density, such as near the bubble walls with $n_{\rm H} \sim 1\,\mathrm{cm^{-3}}$ and a temperature of $T \sim 10^4$\,K, the factor $\dot{a}(T)\,n_{\rm H}$ obtains an order of magnitude of $10^{-10}\,\mathrm{s^{-1}}$.
With the diffusion coefficient of $10^{28}\,\mathrm{cm^2\,s^{-1}}$, the second term in the integral of Eq.\,(\ref{eq:general_solution_deq}) reduces to $\sim$$10^{-38}\,\mathrm{cm^{-2}} \approx 0.1\,\mathrm{pc^{-2}}$.
This means that the propagation is hampered severely beyond the pc scale.
Consequently, we can assume that the annihilation of positrons -- once they are cooled down sufficiently -- is instantaneous.
Finally, assuming a smooth step function (Fermi function) in density and temperature, the typical width of such an instantaneous positron annihilation region is at most 1\,pc and exponentially decreasing.
The exponential decrease depends inversely proportional on $D$ (see Eq.\,(\ref{eq:general_solution_deq})) with $D \sim 10^{28}\,\mathrm{cm^2\,s^{-1}}$ leading to a sharp profile with at most $10^{-3}$\,pc width.
Such a profile will essentially outline the inner boundaries of the geometrical Local Bubble model.

The absolute luminosity of the annihilating positrons is derived from the \nuc{Al}{26} ejecta mass and the delay between the supernova explosion plus the propagation time scale (cooling time),
\begin{equation}
    T_\mrm{cool} = \int_{E_i}^0\frac{\mrm{d}E}{\left| \frac{\mrm{d}E}{\mrm{d}t} \right|}\,,
    \label{eq:cooling_time}
\end{equation}
which evaluates to at most 0.4\,Myr in the case of $n_{\rm H} \sim 10^{-4}\,\mathrm{cm^{-3}}$.
We use this value as the maximum delay time $t_{\rm prop}$ of positrons which are seen to annihilate \emph{now}, so that the annihilation luminosity reads
\begin{equation}
    L_{\pm} \approx \sum_{k=1}^2 \left[\frac{M_{26,k}\,p_{\beta+}}{m_{26}\,\tau_{26}}\,\exp\left(-\frac{t_k}{\tau_{26}}\right) \right] \, \exp\left(+\frac{t_{\rm prop}}{\tau_{26}}\right)\,,
    \label{eq:positron_luminosity_case1}
\end{equation}
where $t_1 = 3$\,Myr and $t_2 = 7$\,Myr are the ages of the two supernovae, and $M_{26,1} = M_{26,2} = (1.6$--$13.0) \times 10^{-5}\,\mathrm{M_\odot}$ are the respective \nuc{Al}{26} yields.
Eq.\,(\ref{eq:positron_luminosity_case1}) then evaluates to $10^{36}$--$10^{37}\,\mathrm{e^+\,s^{-1}}$.
This luminosity will be distributed on a shell about 1\,pc thick, which is shaped as the Local Bubble, resulting in an emissivity of $\sim$$10^{-23}$--$10^{-22}\,\mathrm{e^+\,cm^{-3}\,s^{-1}}$.
We discuss the systematic uncertainties of this model in Sect.\,\ref{sec:discussion}.

\begin{figure}[!ht]
    \centering
    \includegraphics[width=1.0\columnwidth]{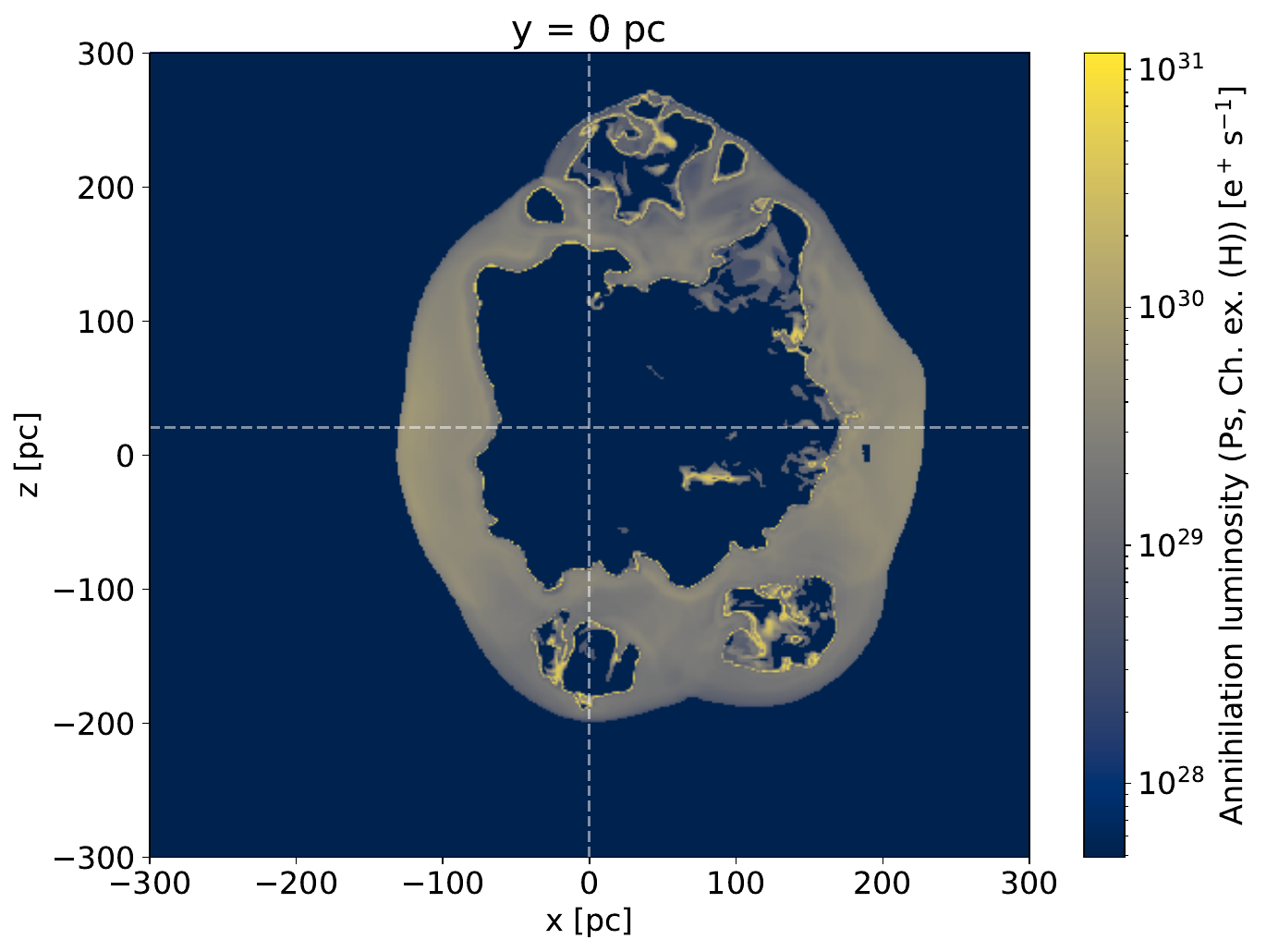}
    \caption{Annihilation luminosity expected in the simulation by \citet{Schulreich2023_LB} from only the charge exchange annihilation channel. Shown is the cell-averaged total luminosity for a slice through $y = 0\,$pc. Clearly, boundary of the supershell shows the highest luminosity, together with the clouds inside the bubble. We again use the condition that the flow velocity must be larger than $1\,\mathrm{km\,s^{-1}}$ to distinguish the bubble from the background medium (see also Fig.\,\ref{fig:mass_distribution_per_nH}).}
    \label{fig:annihilation_luminosity_case2}
\end{figure}

\subsection{Case 2: Hydrodynamics simulations}\label{sec:case2_general}
In order to apply the line-of-sight integration technique described in Sect.\,\ref{sec:case1_general} without major modifications to those snapshots of the hydrodynamics simulations by \citet{Schulreich2023_LB} that capture the Local Bubble at the present time, we
interpolated the values from the adaptively gridded Cartesian mesh onto a uniform Cartesian one having the same resolution as the finest grid refinement level.
As a result, the edge length, $\Delta x$, of each cubic grid cell is about 0.781\,pc, which translates into a cell volume of $\Delta V = (\Delta x)^3 \approx 0.477$\,pc$^3$.
The starting point of the integration in Eq.\,(\ref{eq:los_coordinates}) is now the actual location of the Sun at $\left(x,y,z\right)_\odot \approx (0,0,20.8)$\,pc according to \citet{Bennett2018_Sun_GaiaDR2}, which was also used in the simulations by \citet{Schulreich2023_LB}.
Of course, the observer can, in principle, be placed at any point, even outside the Local Bubble.
We demonstrate this in Sect.\,\ref{sec:outside_view_case2} where we briefly examine what a 10--20\,Myr-old superbubble, reminiscent of the Local Bubble, would look like at a distance of a few 100\,pc.

Before calculating the emissivities for the intermediate-lived radioisotopes and for positron annihilation, we investigate where the two isotopes are found in relation to the overall gas density.
To do so, we derived the cell-averaged hydrogen number density, $n_\mrm{H}(x,y,z)=\rho(x,y,z)\,X/m_\mathrm{p}$, and the mass of isotope $i$, $M_i(x,y,z)=\rho_i(x,y,z)\,\Delta V$, from the simulation data, where $\rho$ is the (total) gas mass density, $X\approx 0.707$ is the hydrogen mass fraction, $m_\mrm{p}$ is the proton mass, and $\rho_i$ is the mass density of isotope $i$.
Binning $n_\mrm{H}$ logarithmically in $1000$ steps between $10^{-5}$ and $10^1\,\mathrm{cm^{-3}}$, we obtain bimodal mass distributions for both radioisotopes peaking around $10^{-4}$ and $10^0\,\mathrm{cm^{-3}}$, which corresponds to the diluted Local Bubble interior and its dense outer shell, respectively.
In regions with gas densities between $10^{-3}$ and $10^{-1}\,\mrm{cm^{-3}}$, significantly fewer radioisotopes are present.
Furthermore, we separately sum the \nuc{Al}{26} and \nuc{Fe}{60} masses found in each hydrogen number density bin.
The resulting distributions and cumulative distributions are shown in Fig.\,\ref{fig:mass_distribution_per_nH}.
It can be seen that most ($\sim$$80\,\%$) of the \nuc{Fe}{60} is found in the denser material ($n_{\rm H} \gtrsim 10^{-1}\,\mrm{cm^{-3}}$), whereas more than $\sim$$50\,\%$ of \nuc{Al}{26} is found in more diluted gas phases ($n_{\rm H} \lesssim 10^{-1}\,\mrm{cm^{-3}}$; see Fig.\,\ref{fig:mass_distribution_per_nH}, right).
Deviations are due to the significantly different decay times of the two radioisotopes, as well as their explosive yields, which not only show strongly divergent dependencies on the initial masses and thus explosion times of the Local Bubble progenitor stars, but are also generally lower for \nuc{Al}{26} than for \nuc{Fe}{60} \citep[see table~2 in][]{Schulreich2023_LB}.
Furthermore, in the model only supernovae contribute to the \nuc{Fe}{60} present today, while for \nuc{Al}{26} also stellar winds have an impact, which only cease before 0.88\,Myr (case 2a) or before 1.68\,Myr (case 2b), the times of the last supernova explosions.
All this is reflected in a relatively balanced distribution between cavity and supershell for \nuc{Al}{26}, whereas \nuc{Fe}{60}, which is generally present in higher masses, is found proportionally more in the supershell. 
As a consequence, the \nuc{Al}{26} and \nuc{Fe}{60} ejecta are not co-spatial, which, however, is a typically assumption made in $\gamma$-ray data analyses \citep[e.g.][]{Wang2007_60Fe,Wang2020_Fe60,Siegert2023_PSYCO}.
We further discuss the implications of this finding in Sect.\,\ref{sec:discussion}.

\subsubsection{\nuc{Fe}{60} and \nuc{Al}{26} emissivities}\label{sec:case2_nuc_emissivities}
We show slices of the \citet{Schulreich2023_LB} hydro-simulation at $y = 0$\,pc for the \nuc{Al}{26} and \nuc{Fe}{60} number densities in Fig.\,\ref{fig:MS_slice_26Al_60Fe}.
Also in this representation it is evident that the distribution between the superbubble interior and shell is more balanced for \nuc{Al}{26} than for \nuc{Fe}{60}, of which relatively more mass could be swept into the shell, primarily by supernova blast waves.
From the masses as a function of Cartesian coordinates, we can straight-forwardly calculate the $\gamma$-ray emissivities (cf. Eq.\,(\ref{eq:emissivity_norm})),
\begin{equation}
    \epsilon_i(x,y,z) = M_i(x,y,z)\,\frac{p_i}{m_i\,\tau_i}\,,
    \label{eq:nuc_emissivities_case2}
\end{equation}
from which the flux as a function of longitude and latitude is calculated by Eq.\,(\ref{eq:los_case1}).

\subsubsection{Positron annihilation from \nuc{Al}{26}}\label{sec:case2_positrons}
Similar to case 1, we will estimate where the positrons from \nuc{Al}{26} annihilate in the Local Bubble, given the temperature and density in the simulation.
The minimum and maximum temperatures in the simulation are 8000\,K and 27.8\,MK, respectively.
The effect of diffusion will be treated in the same way as before, so that we assume instantaneous annihilation in regions with a preferential combination of the annihilation rate and target density.
In Fig.\,\ref{fig:MS_slice_nH_Temp} we show slices at $y=0$ through the simulated hydrogen number density and temperature distribution in the Local Bubble region.
Naturally, one would expect the positrons to annihilate whenever they reach a gas parcel with sufficiently high density and low temperature.
Positrons undergo annihilation with neutral hydrogen via charge exchange, taking the electron from the hydrogen atom, to build Ps.
However, also annihilations with free electrons can occur, depending on the energy of the positrons and the density of electrons.
Since the electron density is not available in this simulation, we restrict ourselves to the case of charge exchange with hydrogen since it is anyway the most probable annihilation channel (see Fig.\,\ref{fig:annihilation_rates}).

At the bubble walls with temperatures around $10^4$\,K, the charge exchange with hydrogen is about two orders of magnitude more probable than the direct annihilation or radiative recombination with electrons.
For higher temperatures, around the peak of the annihilation rates at $2 \times 10^5$\,K, the probability is even $10^5$ times greater for charge exchange.
For very high temperatures, beyond $\sim 2$\,MK, where all hydrogen can be expected to be ionised, the charge exchange naturally drops to zero, and radiative recombination with free electrons would take over as the dominant channel.
However, the annihilation rate at these high temperatures and the available number of electrons to annihilate with (assuming a one-to-one correlation between protons and electrons), will make the contribution to the $\gamma$-ray signal completely negligible.

For the positron emissivity $\epsilon_\pm(x,y,z)$ in units of $\mathrm{e^+\,cm^{-3}\,s^{-1}}$, we calculate 16 logarithmic temperature bins from $8000$\,K to $27.8$\,MK to assign an annihilation rate to each $(x,y,z)$-pixel.
By multiplying with the cell-averaged hydrogen number densities, we can estimate the weights for the positron annihilation rates 
\begin{equation}
    \dot{r}_\pm(x,y,z) = \dot{a}(T(x,y,z))\,n_\mrm{H}(x,y,z)\,,
    \label{eq:annihilation_rates_case2}
\end{equation}
which is shown in Fig.\,\ref{fig:annihilation_luminosity_case2} when converted to a positron luminosity $L_\pm(x,y,z)$.
The exact number of positrons annihilating \emph{now} would, again, require to estimate the \nuc{Al}{26} positron production rate, the propagation time, and annihilation time.
For simplicity, we assume that the \nuc{Al}{26} positron production rate now will result in the positron annihilation rate now, $L_\pm(x,y,z) \leftarrow \dot{r}_\pm(x,y,z)/\sum(\dot{r}_\pm(x,y,z)) L_{\beta+}$, where 
\begin{equation}
    L_{\beta+} = \sum_{x,y,z}\left[\frac{M_{26}(x,y,z)\,p_{\beta+}}{m_{26}\,\tau_{26}}\right] \approx 1.9 \times 10^{37}\,\mrm{e^+\,s^{-1}}\,.
    \label{eq:total_positron_luminosity_case2}
\end{equation}

\begin{figure*}
    \centering
    \includegraphics[width=1.0\columnwidth]{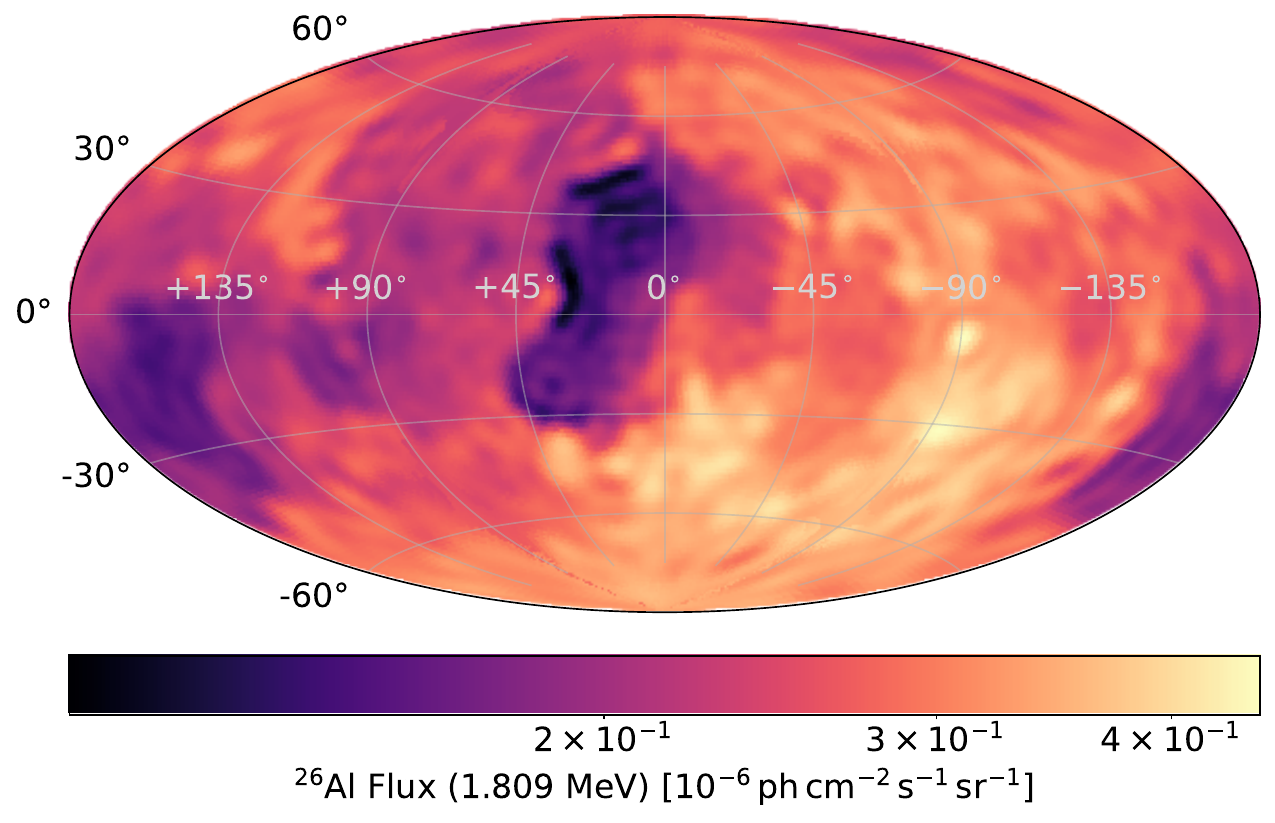}~
    \includegraphics[width=1.0\columnwidth]{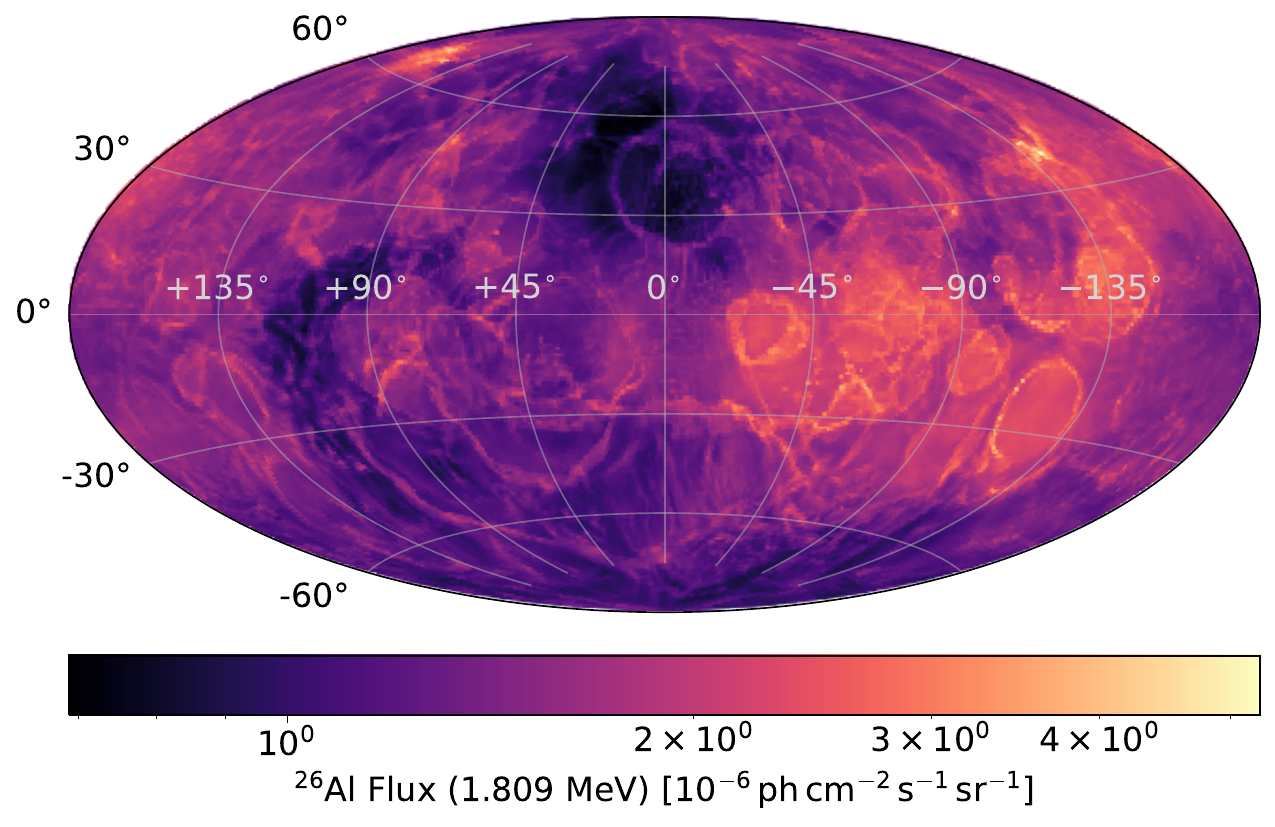}
    \caption{Expected all-sky maps from the decay of \nuc{Al}{26} at 1809\,keV in the Local Bubble. Left: Case 1a of the geometrical model with a total flux of $3.3 \times 10^{-6}\,\mathrm{ph\,cm^{-2}\,s^{-1}}$. Right: Case 2a of the hydrodynamics simulation with a total flux of $19.5 \times 10^{-6}\,\mathrm{ph\,cm^{-2}\,s^{-1}}$. See text for discussion and details.}
    \label{fig:images_26Al}
\end{figure*}
\begin{figure*}
    \centering
    \includegraphics[width=1.0\columnwidth]{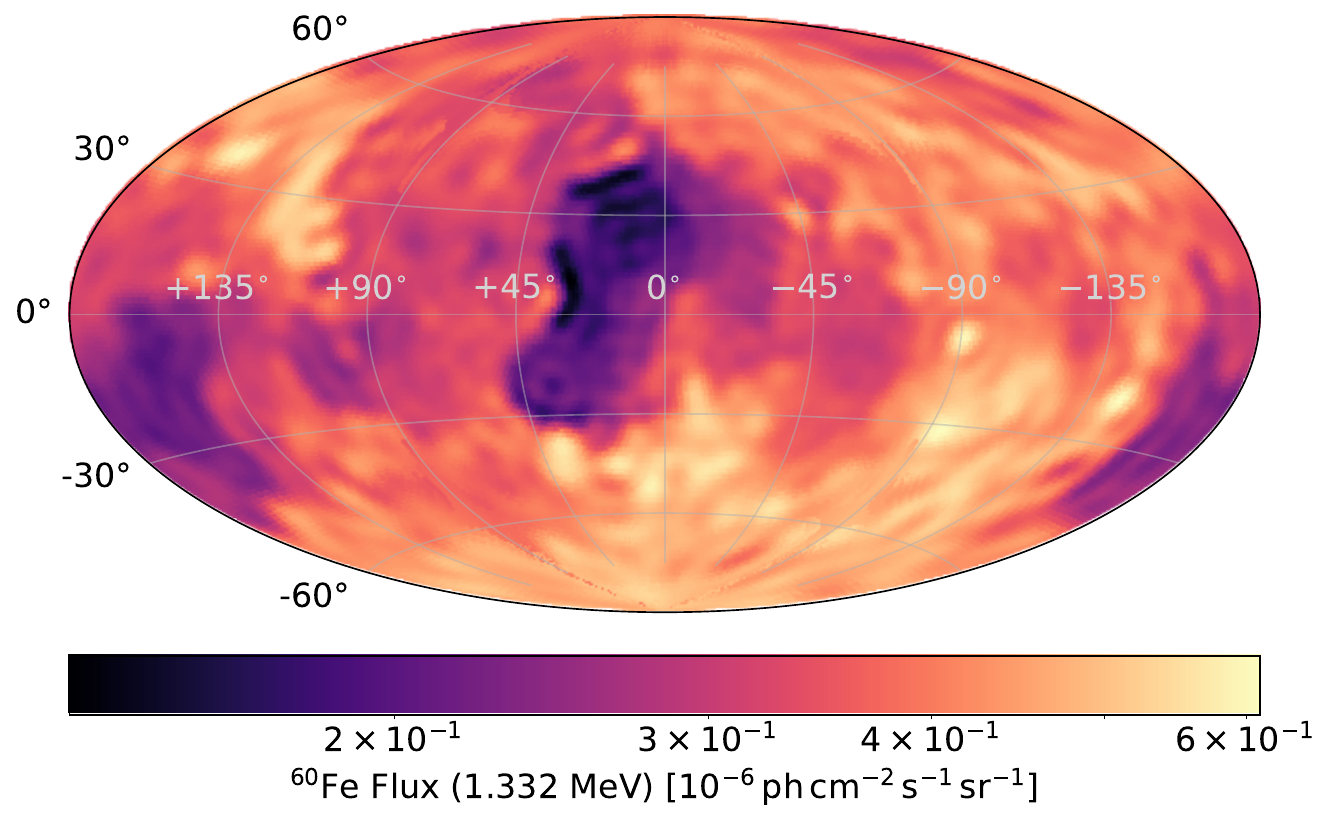}
    \includegraphics[width=1.0\columnwidth]{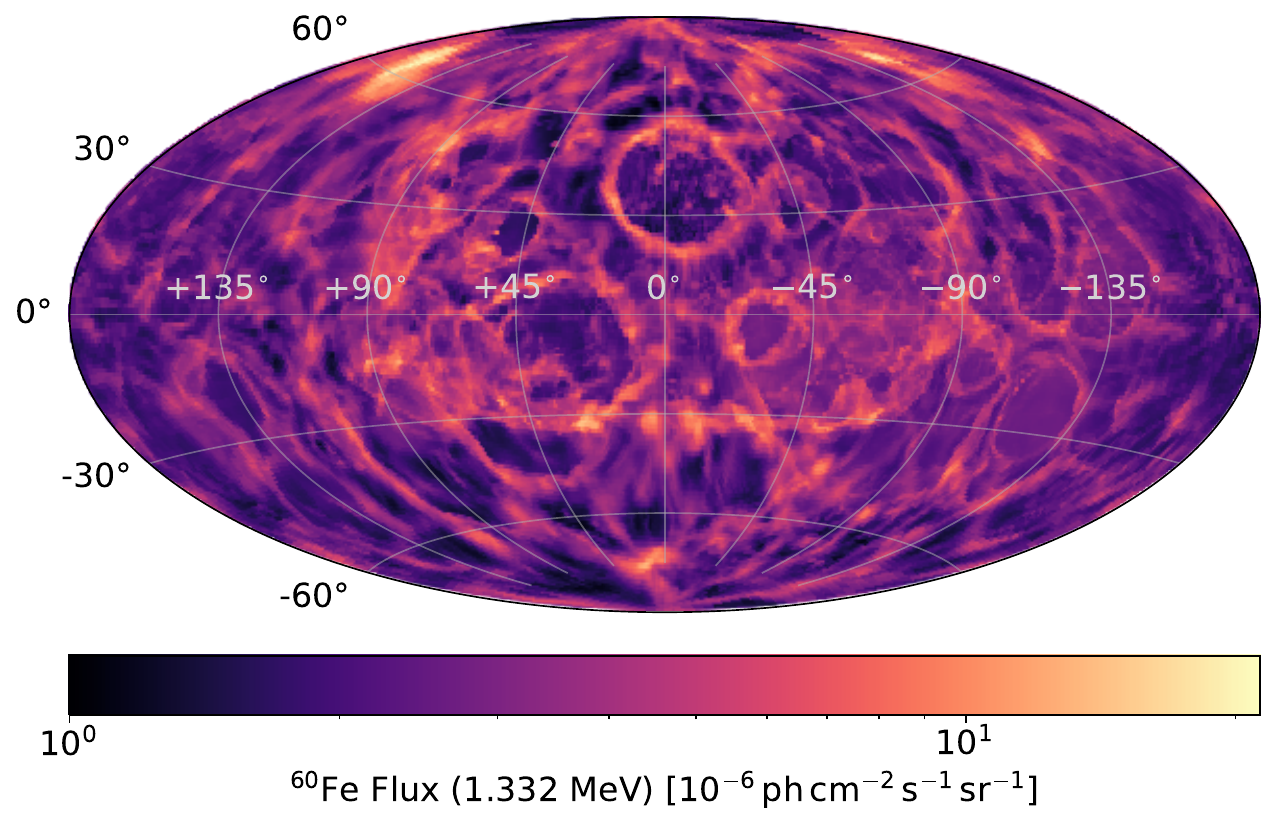}
    \caption{Same as Fig.\,\ref{fig:images_26Al} but for the $\gamma$-ray line at 1332\,keV from \nuc{Fe}{60}. The total fluxes are $4.6 \times 10^{-6}\,\mathrm{ph\,cm^{-2}\,s^{-1}}$ and $42.2 \times 10^{-6}\,\mathrm{ph\,cm^{-2}\,s^{-1}}$, respectively.}
    \label{fig:images_60Fe}
\end{figure*}
\begin{figure*}
    \centering
    \includegraphics[width=1.0\columnwidth]{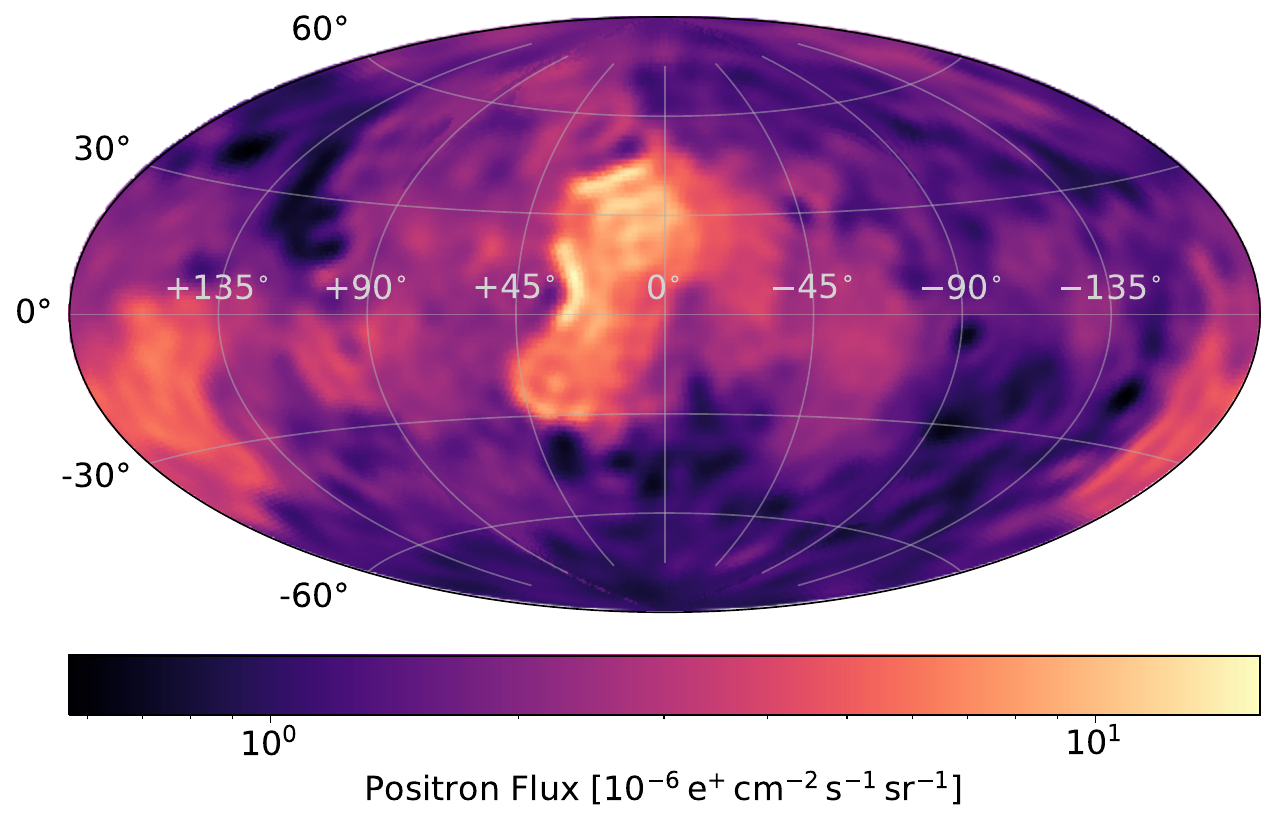}
    \includegraphics[width=1.0\columnwidth]{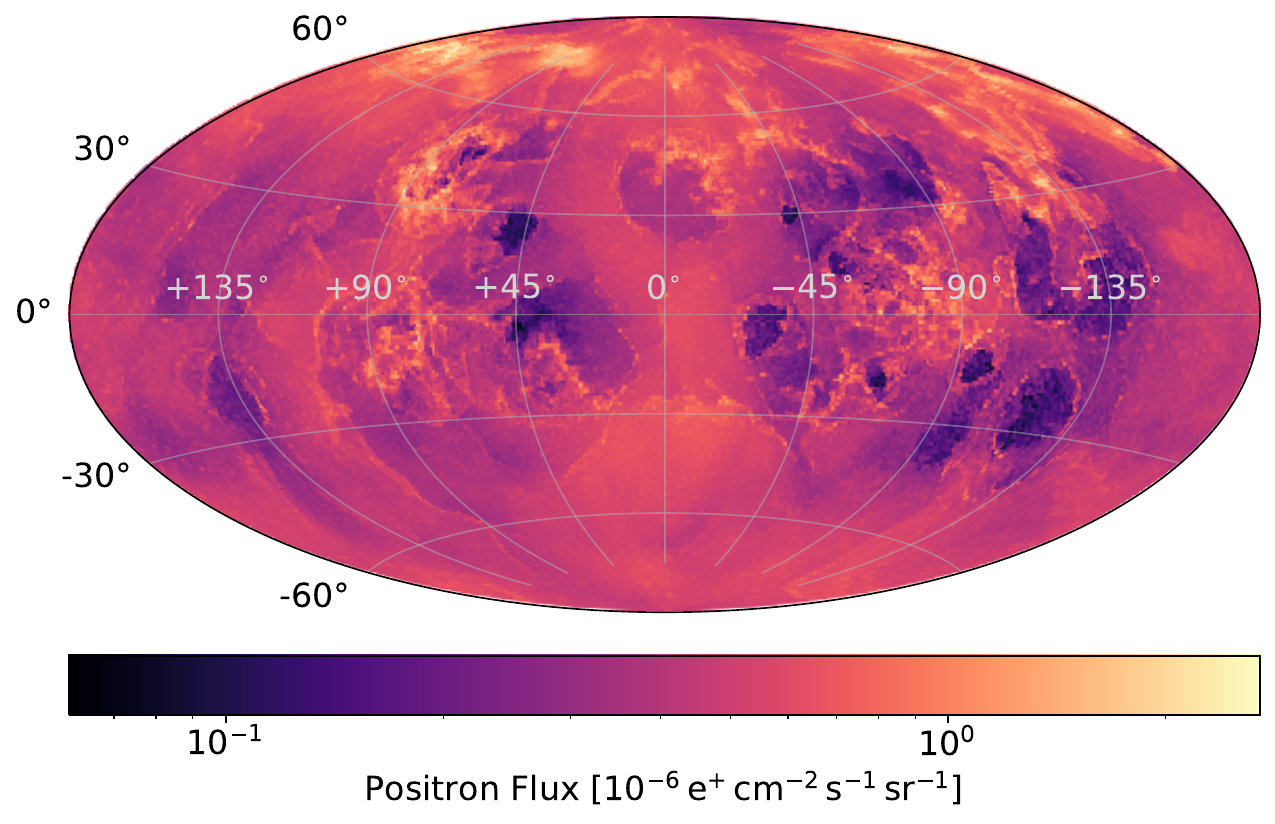}
    \caption{Same as Fig.\,\ref{fig:images_26Al} but for the positron flux. The conversion to 511\,keV flux is a factor of $0.568$ (see text). The total 511\,keV line fluxes are then $21.4 \times 10^{-6}\,\mathrm{ph\,cm^{-2}\,s^{-1}}$ and $2.0 \times 10^{-6}\,\mathrm{ph\,cm^{-2}\,s^{-1}}$, respectively.}
    \label{fig:images_positrons}
\end{figure*}

\section{$\gamma$-ray line images from the Local Bubble}\label{sec:results}
We apply Eq.\,(\ref{eq:los_case1}) for the $\gamma$-ray line of \nuc{Al}{26} at 1808.63\,keV, the line of \nuc{Fe}{60} at 1332.5\,keV, and the positron annihilation line originating only from \nuc{Al}{26} decay at 511\,keV to the two previously discussed cases.
We compare the images of case 1a and case 2a side by side for each line and summarise the properties of the models in Tab.\,\ref{tab:flux_summary}.

\begin{figure*}
    \centering
    \includegraphics[width=1.0\columnwidth]{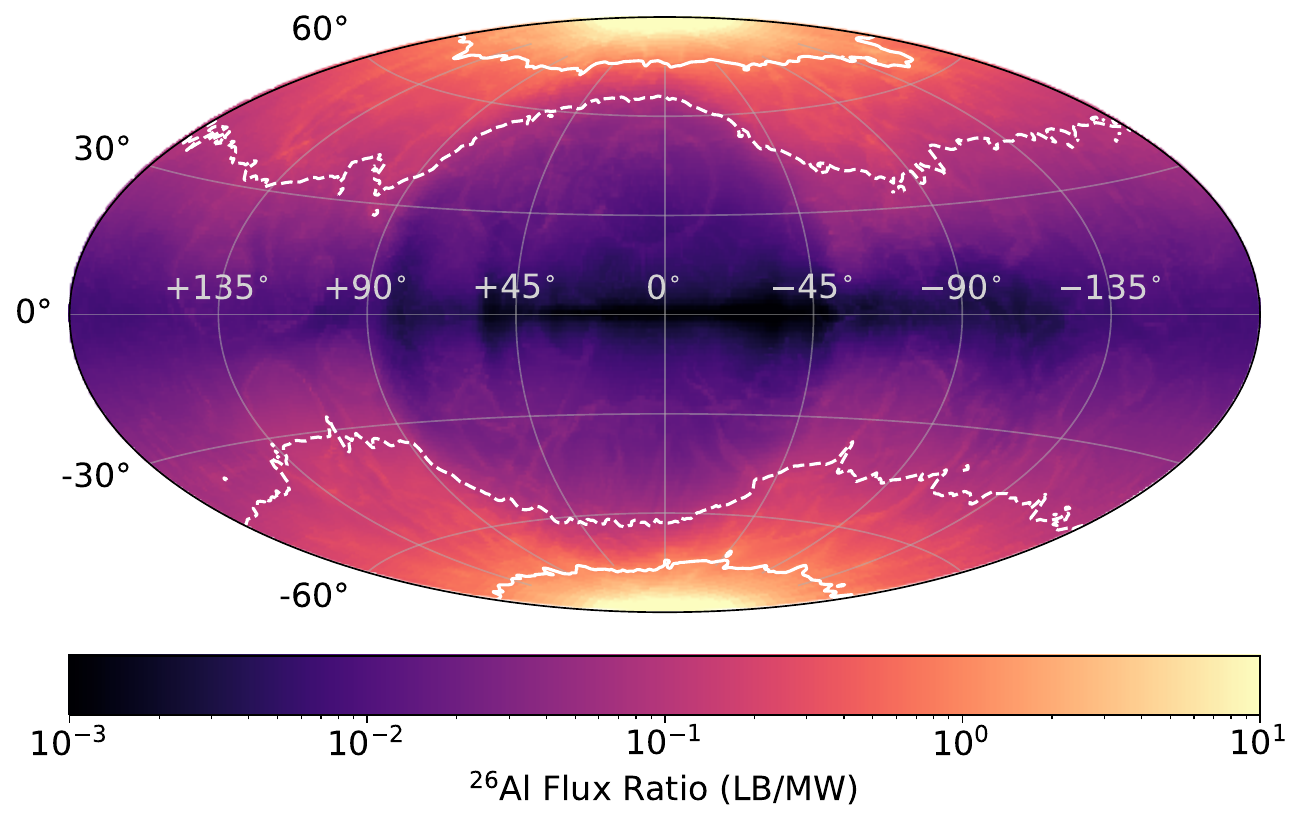}~
    \includegraphics[width=1.0\columnwidth]{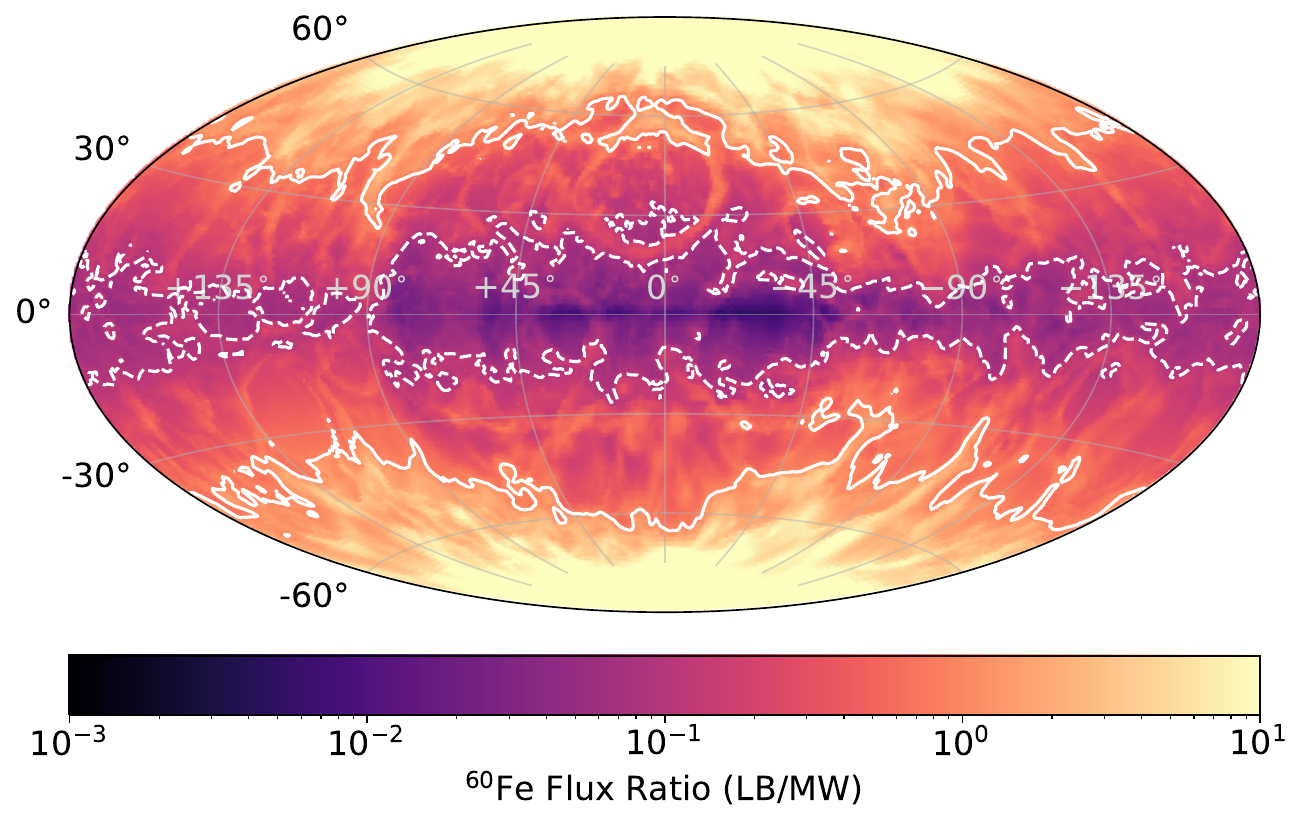}
    \caption{Flux ratios of the Local Bubble (LB) all-sky maps (case 2a) and the Milky Way (MW) model for the \nuc{Al}{26} (left) and \nuc{Fe}{60} lines. The solid (dashed) lines show the regions where the Local Bubble is at least $1.0$ ($0.1$) as strong as the Milky Way. The typical values for when the Local Bubble contribution might be stronger than the Galactic background shown here for \nuc{Al}{26} and \nuc{Fe}{60} are $|b| \gtrsim 73^\circ$ and $|b| \gtrsim 48^\circ$, respectively.}
    \label{fig:relative_maps}
\end{figure*}

\subsection{\nuc{Al}{26} emission at 1809\,keV}\label{sec:Al26_results}
In Fig.\,\ref{fig:images_26Al} we show the comparison of the resulting 1809\,keV $\gamma$-ray line maps from the decay of \nuc{Al}{26} between case 1a with maximal structure (left) and case 2a with 14 supernovae (right).
Clearly, because the spatial resolution of the hydro-simulation is much higher than that of the geometrical model, there are more structures visible in the line-of-sight integrated map of the simulation.
This will be similar throughout the comparisons and we omit repeating this statement in the following.
In the map of case 2a, many of the `fingers' from Fig.\,\ref{fig:MS_slice_26Al_60Fe} are now visible as framed features.
The direction, albeit not the distance, of the last supernova is also evident as the region between $-140^\circ \lesssim \ell \lesssim -20^\circ$ and $-40^\circ \lesssim b \lesssim +30^\circ$ is brighter than the remaining sky.
Due to the \nuc{Al}{26} lifetime of 1.05\,Myr, the contribution from the last supernova, as well as from the progenitor's winds, did not yet homogenise inside the bubble as opposed to the case for \nuc{Fe}{60} (see Fig.\,\ref{fig:images_60Fe}, right).
We mimic this behaviour in the geometrical model (case 1), and placed the more recent of the two supernovae considered (3\,Myr) at $(x,y,z)_{\rm SN} = (40.9,-65.9,19.5)$\,pc (see table~2 in \citet{Schulreich2023_LB}), which leads to brighter emission in the same regions in the left image, compared to the remaining sky.
Likewise, the same effect is visible in the geometric model of \nuc{Fe}{60} (see Fig.\,\ref{fig:images_60Fe}, left), which only appears slightly more homogeneous due to the longer lifetime of \nuc{Fe}{60}.

The total 1809\,keV $\gamma$-ray line fluxes integrated across the entire sky for case 1a and case 2a are $3.3 \times 10^{-6}\,\mathrm{ph\,cm^{-2}\,s^{-1}}$ and $19.5 \times 10^{-6}\,\mathrm{ph\,cm^{-2}\,s^{-1}}$, respectively.
The total isotropic part in each image, calculated by
\begin{equation}
    F_{\rm ISO} = \int d\Omega\, \mrm{min}\left(F(\ell,b)\right)\,,
    \label{eq:isotropic_flux}
\end{equation}
is $1.3 \times 10^{-6}\,\mathrm{ph\,cm^{-2}\,s^{-1}}$ ($39\,\%$) and $8.7 \times 10^{-6}\,\mathrm{ph\,cm^{-2}\,s^{-1}}$ ($44\,\%$), respectively.
Although the total fluxes differ by a factor of $\sim$6, the isotropic fraction of the emission models are similar between 40 and 50\,\%.
Comparing cases 1a and 1b, the fluxes are similar, with the isotropic contribution to be enhanced by 10\,\% due to the smoother reconstruction of the inner radii.
As opposed to case 1, the difference between case 2a and case 2b is almost a factor of three in total flux of the \nuc{Al}{26} line, mainly because the last supernova has been simply taken out.
Then, the isotropic contribution is reduced by $\sim$15\,\% because much of the \nuc{Al}{26} has mixed with the bubble walls.
Clearly, the impact of the most recent supernova is, expectedly, the largest and would gauge all measurements.
The differences and systematic uncertainties in the estimates will be further discussed in Sect.\,\ref{sec:discussion}.

\subsection{\nuc{Fe}{60} emission at 1332\,keV}\label{sec:Fe60_results}
In Fig.\,\ref{fig:images_60Fe} we show the comparison of the resulting 1332\,keV $\gamma$-ray line maps from the decay of \nuc{Fe}{60} between case 1a with maximal structure (left) and case 2a with 14 supernovae (right).
As opposed to the general belief that \nuc{Al}{26} and \nuc{Fe}{60} are co-spatial inside of superbubbles, we clearly find a different appearance between the isotopes.
This is true for both cases, although the hydro-simulations provide a better point of argumentation for \nuc{Fe}{60}.
In the case of \nuc{Fe}{60}, mostly the high-density regions are protruding features, now outlining the boundaries at the superbubble walls.
These ring-like structures are projections of the irregularities of the superbubble, created probably by thermal and hydrodynamic instabilities as well as additional pressure from individual supernovae.
This leads to a more homogeneous appearance, even though individual high density features, such as around $\ell \sim 135^\circ$, $b \sim 65^\circ$ can outshine the remaining sky by factors of a few.
The $\gamma$-ray maps in 1332\,keV show no strong preferred direction of the flux, apart from the different distances to wall edges (left) leading to a contrast of at most $\sim$$6$, and apart from high density regions which may be prone to forming new stars (right), leading to a contrast of at most $\sim$$15$.

The total 1332\,keV $\gamma$-ray line fluxes integrated across the entire sky for case 1a and case 2a are $4.6 \times 10^{-6}\,\mathrm{ph\,cm^{-2}\,s^{-1}}$ and $42.2 \times 10^{-6}\,\mathrm{ph\,cm^{-2}\,s^{-1}}$, respectively.
The total isotropic part in each image is $1.7 \times 10^{-6}\,\mathrm{ph\,cm^{-2}\,s^{-1}}$ ($36\,\%$) and $13.3 \times 10^{-6}\,\mathrm{ph\,cm^{-2}\,s^{-1}}$ ($31\,\%$), respectively.
Again, the total fluxes differ by a factor of $\sim 9$ but the isotropic fraction of the emission models are similar between 30 and 40\,\%.
Comparing case 1a to 1b leads again to a similar total flux and slightly increased isotropic ratio as the boundaries are smoother.
Likewise, case 2b is showing a $\sim 25$\,\% lower flux than case 2a owing to the removal of the last supernova.
This reduction in \nuc{Fe}{60} line flux is not as strong as for \nuc{Al}{26} because of the longer lifetime of \nuc{Fe}{60}.
More differences and systematic uncertainties will be further discussed in Sect.\,\ref{sec:discussion}.

\subsection{Positron annihilation emission}\label{sec:positron_results}
The positron annihilation images in Fig.\,\ref{fig:images_positrons} are almost similar to the `negatives' of the \nuc{Al}{26} images (Fig.\,\ref{fig:images_26Al}).
This is expected from the distance to the bubble walls where most of the positron annihilation should happen:
a bubble wall that is farther away leads to more \nuc{Al}{26} along the line of sight to integrate over and consequently a larger 1809\,keV flux, but at the same time to a smaller positron annihilation flux as it scales per inverse distance squared.
In case 1, this is clearer visible compared to case 2 with the large nearby feature around $-5^\circ \lesssim \ell \lesssim +45^\circ$ and $-30^\circ \lesssim b \lesssim +45^\circ$. 
The positron annihilation image of case 2a appears more homogeneous than the \nuc{Al}{26} map.
This is also understandable as the simulation is bound by a large density in the Galactic plane (see Fig.\,\ref{fig:mass_distribution_per_nH}, left).
We set a limit to the annihilation belonging to the Local Bubble wherever the flow velocity is above $1\,\mathrm{km\,s^{-1}}$.
In doing so, we set a physically-motivated boundary for where we assume positron annihilation to happen if it originated from only \nuc{Al}{26} inside the Local Bubble.
Clearly, the details in the resulting maps of case 2 originate in the high spatial resolution of the hydrodynamics simulation.

In Fig.\,\ref{fig:images_positrons}, we show the positron flux that will undergo charge exchange with hydrogen, not the positron annihilation flux directly.
By charge exchange, one positron will capture one electron to form Ps.
Depending on the spin state, Ps will either decay into two photons yielding a 511\,keV line (para-Ps), or into three photons yielding a rising continuum up to 511\,keV (ortho-Ps) \citep[][see also Eqs.\,(\ref{eq:pPs-decay}--\ref{eq:oPs-decay})]{Ore1949_511}.
The maximum quantum-statistical ratio between the luminosities (and therefore fluxes) of ortho- and para-Ps is $4.5:1$, as para-Ps is formed $1/4$ of the time resulting in 2 photons, and ortho-Ps $3/4$ of the time resulting in 3 photons.
The total Ps decay flux (para-Ps and 511\,keV line direct plus ortho-Ps) is therefore calculated from the positron flux as
\begin{equation}
    F_{\pm} = F_{\rm 511} + F_{\rm oPs} = \left[2\,\frac{1}{4}\,f_{\rm Ps} + 2\,(1-f_{\rm Ps}) \right] F_{\rm e^+} + 3\, \frac{3}{4}\,f_{\rm Ps}\, F_{\rm e^+}\,,
    \label{eq:Ps_flux_total}
\end{equation}
from which the para-Ps (plus 511\,keV line) and ortho-Ps flux (continuum) contributions are readily seen \citep{Brown1987_511}.
Using the above definition includes the Ps fraction $f_{\rm Ps}$, which is expected to be $0.955$ for a completely neutral hydrogen medium \citep{Guessoum2005_511}.
Considering only the 511\,keV line, we obtain a conversion of $F_{\rm 511} = 0.568 F_{\rm e^+}$\footnote{The conversion for the ortho-Ps flux would then be $F_{\rm oPs} = 2.149 F_{\rm e^+}$}.

The total 511\,keV $\gamma$-ray line fluxes integrated across the entire sky for case 1a and case 2a are then $21.4 \times 10^{-6}\,\mathrm{ph\,cm^{-2}\,s^{-1}}$ and $2.0 \times 10^{-6}\,\mathrm{ph\,cm^{-2}\,s^{-1}}$, respectively.
The total isotropic part in each image is $5.3 \times 10^{-6}\,\mathrm{ph\,cm^{-2}\,s^{-1}}$ ($25\,\%$) and $0.2 \times 10^{-6}\,\mathrm{ph\,cm^{-2}\,s^{-1}}$ ($13\,\%$), respectively.
The total fluxes differ by a factor of $\sim$11, and the isotropic fraction of the emission models are again similar, now being between 15 and 25\,\%.
The differences between case 1a and 1b are, again, similar to the previous $\gamma$-ray lines, with a $\sim$15\,\% increase in the isotropic ratio due to the smoother Local Bubble boundaries.
Surprisingly, cases 2a and 2b are almost identical although the \nuc{Al}{26} input from the last supernova is missing in case 2b.
This may be understood as a geometrical effect as the largest parts of the annihilation flux originate in the thick bubble walls rather than in the first few pc (see Fig.\,\ref{fig:annihilation_luminosity_case2}).
Further differences and systematic uncertainties will be discussed in Sect.\,\ref{sec:discussion}.

\begin{table}
    \centering
    \begin{tabular}{c|rrrr}
         Model & Line  & Total & Isotropic & Isotro- \\
               & energy & flux & flux & pic ratio \\
               & [keV] & [$\mathrm{ph\,cm^{-2}\,s^{-1}}$] & [$\mathrm{ph\,cm^{-2}\,s^{-1}}$] & [\%] \\
        \hline
        LB 1a & $1809$ &  $3.3 \times 10^{-6}$ &  $1.3 \times 10^{-6}$ & $39$ \\
        LB 1b & $1809$ &  $3.4 \times 10^{-6}$ &  $1.6 \times 10^{-6}$ & $48$ \\
        LB 2a & $1809$ & $19.5 \times 10^{-6}$ &  $8.7 \times 10^{-6}$ & $44$ \\
        LB 2b & $1809$ &  $6.7 \times 10^{-6}$ &  $1.9 \times 10^{-6}$ & $28$ \\
        \hline
        LB 1a & $1332$ &  $4.6 \times 10^{-6}$ &  $1.7 \times 10^{-6}$ & $36$ \\
        LB 1b & $1332$ &  $4.7 \times 10^{-6}$ &  $2.1 \times 10^{-6}$ & $43$ \\
        LB 2a & $1332$ & $42.2 \times 10^{-6}$ & $13.3 \times 10^{-6}$ & $31$ \\
        LB 2b & $1332$ & $32.2 \times 10^{-6}$ &  $6.2 \times 10^{-6}$ & $19$ \\
        \hline
        LB 1a &  $511$ & $21.4 \times 10^{-6}$ &  $5.3 \times 10^{-6}$ & $25$ \\
        LB 1b &  $511$ & $20.6 \times 10^{-6}$ &  $7.9 \times 10^{-6}$ & $39$ \\
        LB 2a &  $511$ &  $2.0 \times 10^{-6}$ &  $0.2 \times 10^{-6}$ & $13$ \\
        LB 2b &  $511$ &  $1.7 \times 10^{-6}$ &  $0.2 \times 10^{-6}$ & $12$ \\
        \hline
        \hline
    \end{tabular}
    \caption{Summary of $\gamma$-ray line fluxes for the two model assumptions of the Local Bubble, each with two different model-specific changes. 1) Geometric model: a) $l_{\rm max} = 40$, b) $l_{\rm max} = 6$ (see Fig.\,\ref{fig:inner_radii}). 2) Hydrodynamics simulation: a) 14 SNe, b) 13 SNe. The four cases studied here provide a reasonable range of plausible fluxes. For cases 1a and 1b, we assume ejecta masses of $10^{-4}\,\mathrm{M_\odot}$ for both isotopes (see Sect.\,\ref{sec:case1_nuc_emissivities} for details about the ejecta masses). In cases 2a and 2b, on the other hand, the ejecta mass is the difference between the instantaneous mass of the star at the last data point of the (rotating) stellar evolution model used and its predicted remnant mass, and thus depends on the initial mass of the star. Further systematic uncertainties are discussed in Sect.\,\ref{sec:discussion}.}
    \label{tab:flux_summary}
\end{table}

\subsection{Total $\gamma$-ray line emission: astrophysical background and Local Bubble foreground}\label{sec:total_expectations}
The total $\gamma$-ray line fluxes, integrated over the whole sky, of the Local Bubble are summarised in Tab.\,\ref{tab:flux_summary}.
Compared to the Milky Way emission in the three $\gamma$-ray lines, the Local Bubble flux contribution is marginal, but not negligible.
In \citet{Pleintinger2023_26Al}, the total \nuc{Al}{26} 1809\,keV line flux has been measured as $F_{26}^{\rm MW} = (1.84 \pm 0.03) \times 10^{-3}\,\mathrm{ph\,cm^{-2}\,s^{-1}}$, that is, about a factor of $90$--$600$ times brighter than the Local Bubble alone.
The total \nuc{Fe}{60} flux at 1332\,keV has been measured by \citet{Wang2020_Fe60} to be around $F_{60}^{\rm MW} = (0.31 \pm 0.06) \times 10^{-3}\,\mathrm{ph\,cm^{-2}\,s^{-1}}$, which is about $7$--$70$ times brighter than the Local Bubble.
It is clear from these numbers alone that detecting the Local Bubble with future $\gamma$-ray telescopes (see Sect.\,\ref{sec:COSI_sims}) will require a careful treatment.
However, the almost-isotropic part of this nearby emission as well as the (relatively) stronger flux at very high latitudes ($|b| \gtrsim 60^\circ$) from the Local Bubble compared to the Milky Way (mostly confined to $|b| \lesssim 60^\circ$, see \citet{Pleintinger2019_26Al,Krause2021_26Alchimneys,Siegert2023_PSYCO}) provides some leverage to distinguish fore- and background emission.

We use the NE2001 model by \citet{Cordes2002_NE2001}, as it has been shown to provide a good first-order tracer of the true \nuc{Al}{26} (and potentially also \nuc{Fe}{60}) distribution, to estimate the contribution at different latitudes of the Milky Way and the Local Bubble.
For a fair comparison, we subtract the isotropic part of the NE2001 map as this would represent the foreground emission.
We use the above-mentioned measured flux values for the 1809 and 1332\,keV lines to compare the images.
We find that, indeed, the Local Bubble outshines the Galactic background only for latitudes $|b| \gtrsim 82^\circ$ (case 1) or $|b| \gtrsim 73^\circ$ (case 2) in the case of \nuc{Al}{26}, and $|b| \gtrsim 79^\circ$ (case 1) or $|b| \gtrsim 48^\circ$ (case 2) in the case of \nuc{Fe}{60}.
We show the relative maps for the more optimistic case 2 (higher Local Bubble fluxes) in the 1809 and 1332\,keV lines in Fig.\,\ref{fig:relative_maps}.

\begin{figure}
    \centering
    \includegraphics[width=1.0\columnwidth]{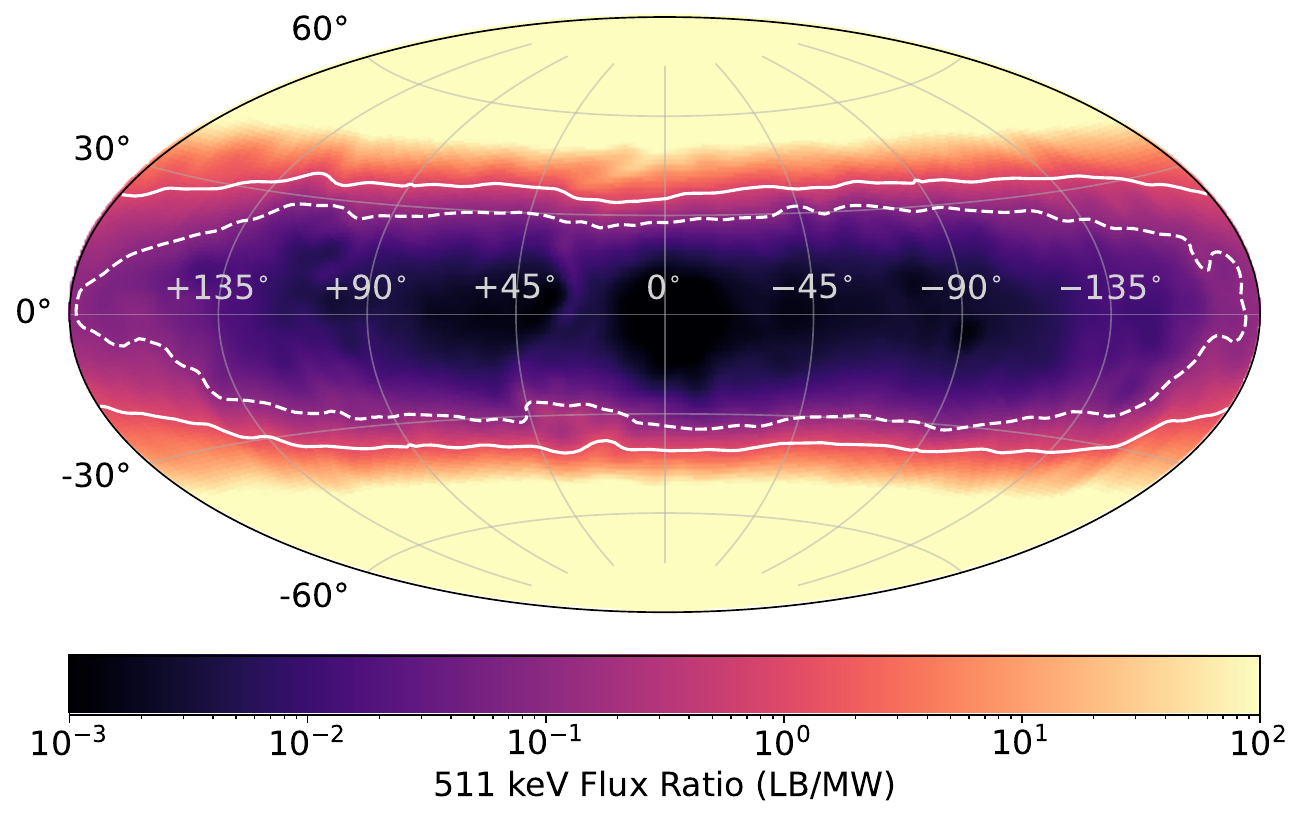}
    \caption{Same as Fig.\,\ref{fig:relative_maps} but for the 511\,keV line in case 1. The figure appears similar for case 2. For latitudes $|b| \gtrsim 36^\circ$, the Local Bubble would dominate the Milky Way background as the flux ratio between the components is larger than 1.}
    \label{fig:relative_511_maps}
\end{figure}

In the case of positron annihilation, and in particular for the 511\,keV line, there is no unique, `good', tracer of the emission that fits the $\gamma$-ray data adequately beyond the empirical models.
The only maps that show a particularly high likelihood are infrared maps between 1.25 and 4.9\,$\mathrm{\mu m}$ \citep{Knoedlseder2005_511,Siegert2023_511}.
In order to provide a fair comparison, we use the four component model of \citet{Siegert2016_511} that includes three 2D-Gaussians to represent the bulge emission and one elongated 2D-Gaussian for the disk which appears to be thick in this analysis ($\sim$2\,kpc).
We note that \citet{Skinner2014_511} provided a similar model but with a thin disk ($\lesssim$300\,pc); we will use the thick disk in the following comparison.
For case 1, the Local Bubble 511\,keV emission would outshine the Galaxy above latitudes $|b| \gtrsim 36^\circ$, which is equivalent to about three standard deviations of the latitudinal width of the Galactic model, $\sigma_{\rm Disk} = 10.5^\circ$.
In case 2, the 511\,keV line of the Local Bubble is weaker, so that only above latitudes $|b| \gtrsim 40^\circ$, the foreground emission would dominate.
We show the relative map in the 511\,keV line for case 1a in Fig.\,\ref{fig:relative_511_maps}.
The respective visualisation for case 2 appears very similar.

Typically, $\gamma$-ray line images in the MeV band are reconstructed in a narrow energy bin around the line of interest.
This means that a part of the diffuse continuum emission underneath the line is also imaged.
Such an additional contribution becomes important, when the spectral resolution of the instrument is poor, so that large parts of the continuum will also be imaged simultaneously to the line, or if the continuum at the respective energy is particularly strong.
For $\gamma$-ray spectrometers such as INTEGRAL/SPI or COSI-SMEX, with high-purity germanium detectors, the resolution is on the order of 1\% or better, so that the continuum emission right at the line energies is typically small.
In the case of poor spectral resolution, such as in the case of most scintillator crystals with 10\% FWHM, for example, this problem will lead to a strong bias in the resulting images and astrophysical interpretation.
Using the works by \citet{Wang2020_Fe60} and \citet{Siegert2022_MWdiffuse}, we estimate the contributions of the diffuse Galactic continuum below the $\gamma$-ray lines at 511, 1173, 1332, and 1809\,keV, respectively, in symmetric bands around the lines according to typical resolutions of germanium detectors (here chosen as 2.1, 2.8, 2.9, and 3.2\,keV FWHM, respectively).
We find that fractions of 11, 45, 39, and 5\%, respectively, of the flux in those images would originate from the diffuse Galactic continuum.
It should be noted that modern image reconstruction algorithms can distinguish between physical processes, and therefore between line and continuum emission, as the full imaging response function can now be taken into account for a spatial-spectral deconvolution.
This means that the estimated contributions from the Galactic diffuse continuum emission will serve as the most conservative cases in the following (see Sect.\,\ref{sec:COSI_sims}).

We note that in all three $\gamma$-ray lines, the isotropic contribution from the Local Bubble cannot be measured with current coded mask telescopes, such as INTEGRAL/SPI \citep[cf.][]{Siegert2022_gammaraytelescopes}.
Only with a sensitive, wide field-of-view, Compton telescope can the contributions from the Galactic background and the Local Bubble foreground be distinguished.
In the next section we evaluate how significant this foreground emission would be for the current design of the future COSI-SMEX mission \citep{Tomsick2023_COSI}.

\begin{table*}
    \centering
    \begin{tabular}{c|r|ccrrr}
        Model & Line energy & Full-sky rate & Total counts in 2\,yr & $\mathrm{Model\,vs.\,MW}$ & $\mathrm{Model\,vs.\,[MW+BG]}$ & Time to reach $5\sigma$ \\
              &  [keV] & [$\mathrm{cnts\,yr^{-1}}$] &  & [$\sigma\sqrt{T/\mathrm{yr}}$] & [$\sigma\sqrt{T/\mathrm{yr}}$] & [$\mathrm{yr}$]\\        
        \hline
        LB 1a & 1809 & $5.8 \times 10^3$ & $1.2 \times 10^4$ & $3.2$ & $2.5$ & $3.9$ \\
        LB 1b & 1809 & $6.0 \times 10^3$ & $1.2 \times 10^4$ & $3.3$ & $2.5$ & $3.8$ \\
        LB 2a & 1809 & $3.4 \times 10^4$ & $6.8 \times 10^4$ & $18.6$ & $14.7$ & $0.1$ \\
        LB 2b & 1809 & $1.2 \times 10^4$ & $2.4 \times 10^4$ & $6.5$ & $5.1$ & $1.0$ \\
        MW    & 1809 & $3.3 \times 10^6$ & $6.6 \times 10^6$ & - & $\gtrsim 1000$ & $2 \times 10^{-5}$ \\
        \hline
        LB 1a & 1332 & $8.7 \times 10^3$ & $1.7 \times 10^4$ & $9.6$ & $4.3$ & $1.4$ \\
        LB 1b & 1332 & $9.0 \times 10^3$ & $1.8 \times 10^4$ & $9.9$ & $4.4$ & $1.3$ \\
        LB 2a & 1332 & $8.0 \times 10^4$ & $1.6 \times 10^5$ & $84.1$ & $38.6$ & $2 \times 10^{-2}$ \\
        LB 2b & 1332 & $6.1 \times 10^4$ & $1.2 \times 10^5$ & $65.0$ & $29.6$ & $3 \times 10^{-2}$ \\
        MW    & 1332 & $8.2 \times 10^5$ & $1.6 \times 10^6$ & - & $\gtrsim 300$ & $2 \times 10^{-4}$ \\
        \hline
        LB 1a & 511 & $4.1 \times 10^4$ & $8.1 \times 10^4$ & $16.9$ & $8.6$ & $0.3$ \\
        LB 1b & 511 & $3.9 \times 10^4$ & $7.8 \times 10^4$ & $16.2$ & $8.3$ & $0.4$ \\
        LB 2a & 511 & $3.6 \times 10^3$ & $7.3 \times 10^3$ & $1.5$ & $0.8$ & $42$ \\
        LB 2b & 511 & $3.2 \times 10^3$ & $6.3 \times 10^3$ & $1.3$ & $0.7$ & $55$ \\
        MW    & 511 & $5.8 \times 10^6$ & $1.2 \times 10^7$ & - & $\gtrsim 1000$ & $2 \times 10^{-5}$ \\
        \hline
        \hline
    \end{tabular}
    \caption{Expectations for COSI-SMEX. From left to right the columns show the model, with cases 1a to 2b (see Tab.\,\ref{tab:flux_summary}), the line energy, the rate expected for each model per year, the total number of photons in the nominal mission time of 2\,yr, the significance of the model vs. the Milky Way (MW) background only (assuming no instrumental background) in units of $\sigma$ scaled per year, the significance of the model vs. the Milky Way and instrumental background together in units of $\sigma$ scaled per year, and the time to reach a significance of $5\sigma$ for each model. In each case, the Milky Way foreground is assumed to have contributions from both the $\gamma$-ray lines and the Galactic diffuse continuum beneath the lines.}
    \label{tab:COSI_values}
\end{table*}

\section{Expectations for COSI-SMEX}\label{sec:COSI_sims}
The Compton Spectrometer and Imager (COSI) is a small explorer (SMEX) mission planned for launch in 2027 \citep{Tomsick2019_COSI,Tomsick2021_COSI,Tomsick2023_COSI}.
COSI-SMEX will be sensitive to soft $\gamma$-ray photons in the energy range between 0.2 and 5.0\,MeV with a spectral resolution from 0.2--1.0\,\% thanks to its Ge detectors.
Due to its wide field of view of $\sim$25\,\% of the sky and its observation strategy that scans the whole sky within two of its low-Earth orbits, COSI-SMEX will be more sensitive to diffuse emission than any of its predecessors owing to the total grasp (effective area times field of view).
The telescope will have an angular resolution on the degree scale which will make it suitable to detect a large number of point sources and extended emission such as from the Local Bubble.
The satellite mission has a nominal duration of 2\,yr for which we normalise the expectations in the following.

We use the latest mass model and corresponding response from COSI's Data Challenge 2 \citep{Karwin2022_COSI_DC} to obtain exposure maps at the required $\gamma$-ray line energies from the dedicated COSI analysis software, \texttt{cosipy} \citep{MartinezCastellanos2023_cosipy}.
The instrumental background is provided by the COSI team (Gallego \& Karwin, priv. comm.) and includes Earth albedo photons, atmospheric neutrons, primary alpha particles, electrons and positrons, primary and secondary protons, and an estimate of the Cosmic Gamma-ray Background (CGB).
By multiplying our calculated model maps with the exposure maps in the respective bands and integrating over the whole sky, we can estimate the number of counts COSI-SMEX would receive per time.
In Tab.\,\ref{tab:COSI_values}, we list the expected rates for the different model assumptions detailed above.

For all three $\gamma$-ray lines considered here, two model assumptions of the Local Bubble (cases 1 and 2; LB 1 and LB 2 in Tab.\,\ref{tab:COSI_values}), and an estimate of the Milky Way (MW), we calculate the count rate COSI-SMEX would receive per year, $R_i$, and the resulting total number of counts in its nominal 2\,yr mission duration, $N_i^{\rm tot}$.
From a simple consideration of the counts per time, we estimate the significance of each component against the Milky Way background only, or against the Milky Way background ($\gamma$-ray line plus diffuse Galactic continuum (cf. Sect.\,\ref{sec:total_expectations}) plus the instrumental background.
For simplicity, we assume that the images of the Galactic diffuse continuum roughly follow the $\gamma$-ray line images at their respective energies.
It is clear from observations \citep[e.g.,][]{Siegert2022_MWdiffuse} that the true spatial distribution from the Galactic diffuse continuum, mostly stemming from Inverse Compton scattering of electrons off the interstellar radiation fields, is not well determined but shows similarities to a normal galaxy with a bulge and a disk.
The significance is therefore calculated by
\begin{equation}
    S_i = \frac{R_i}{\sqrt{R_i + R_{\rm MW} + R_{\rm BG}}}\,,
    \label{eq:signi_estimate}
\end{equation}
where the unit of $S_i$ is the `number of sigmas' scaled per time, i.e. $\sigma\sqrt{T/\mathrm{yr}}$, and can therefore easily be scaled by more or less exposure.
We strongly emphasise that these estimates are bound to uncertainties because Eq.\,(\ref{eq:signi_estimate}) is not strictly applicable, and, especially for small count rates (small significances), wrong \citep[cf.][]{Li1983_LiMa,Vianello2018_sigma}.
A proper estimate would require a complete simulation of all components and a maximum likelihood analysis to ensure the reliability of such significance estimates.
This is beyond the purpose of this study and we refer to Tab.\,\ref{tab:COSI_values} with caution.

The expected all-sky fluxes of the three $\gamma$-ray lines have a strong model dependence (see Tab.\,\ref{tab:flux_summary}):
The geometric model, case 1, typically shows smaller values for nuclear lines of \nuc{Al}{26} and \nuc{Fe}{60}, but a larger value for the 511\,keV line from positron annihilation.
While case 1 is a simple geometric model with heuristic assumptions, case 2 is a detailed hydrodynamics model to match ocean crust measurements, among others.
We therefore created a reasonable bracket for further discussions (Sect.\,\ref{sec:discussion}).
Taking the fluxes at face value, we find that all three lines from the Local Bubble are within reach for COSI-SMEX, either within its nominal mission duration of 2\,yr (most optimistic case), or within some additional years of observations.
In particular, the 1809\,keV line from the Local Bubble may reach a significance of $5\sigma$ above the Milky Way and instrumental background within $\sim$4\,yr -- a $3\sigma$ hint may even be possible in 2\,yr for the least optimistic case.
The \nuc{Fe}{60} lines may, in fact, be easier to detect as the fluxes are typically expected to be larger than the \nuc{Al}{26} line.
We note that the instrumental background around the \nuc{Fe}{60} lines is typically larger, with chances of radioactive build-up on the timescale of years \citep{Diehl2018_BGRDB,Siegert2019_SPIBG}.
In addition will the \nuc{Fe}{60} lines sit on a much stronger diffuse $\gamma$-ray continuum compared to the \nuc{Al}{26} line \citep{Wang2020_Fe60}, which will probably lower the significance estimates in the case of \nuc{Fe}{60}.
The lines at 1332 and 1173\,keV (same branching ratio) may be detected as almost-isotropic contribution within the first two years of COSI-SMEX.
For positron annihilation, the model expectations of case 1 and 2 differ by a factor of $\sim$11, making it either easy to detect such an isotropic contribution within one year, or very difficult to see an isotropic contribution at all from the Local Bubble.
We note that the true emission morphology of positron annihilation from only the Local Bubble may be difficult to disentangle as positrons from other sources, also in addition to \nuc{Al}{26}, may propagate into the Local Bubble and annihilate in its walls.
%

\section{Discussion}\label{sec:discussion}
\subsection{Systematic uncertainties}\label{sec:systmatics}
The calculated flux values for the three $\gamma$-ray lines at 1809, 1332, and 511\,keV from the Local Bubble (Tab.\,\ref{tab:flux_summary}) are subject to uncertainties from the modelling assumptions in each case discussed above.
We estimate the impact of different parameters for the two cases and three lines in the following.

\begin{figure*}[!ht]
    \centering
    \includegraphics[width=1.0\columnwidth]{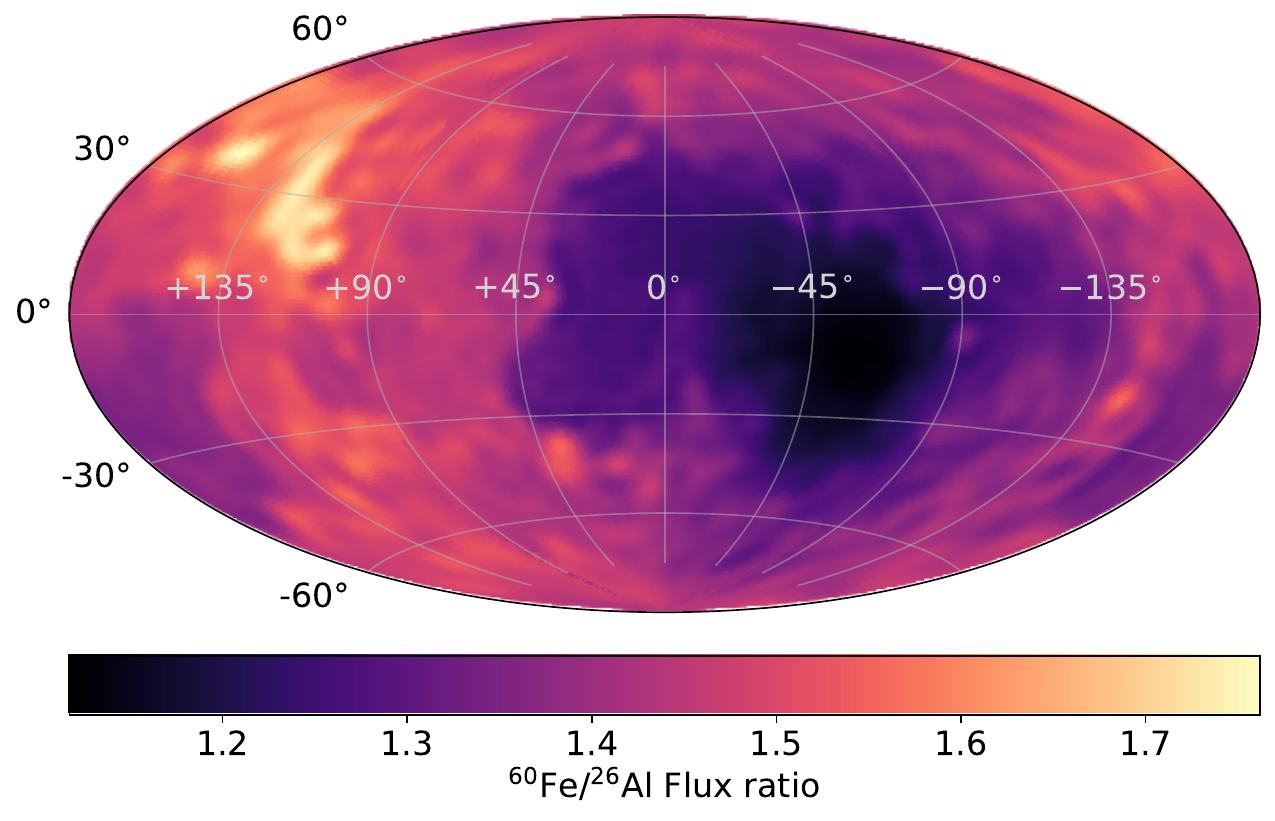}
    \includegraphics[width=1.0\columnwidth]{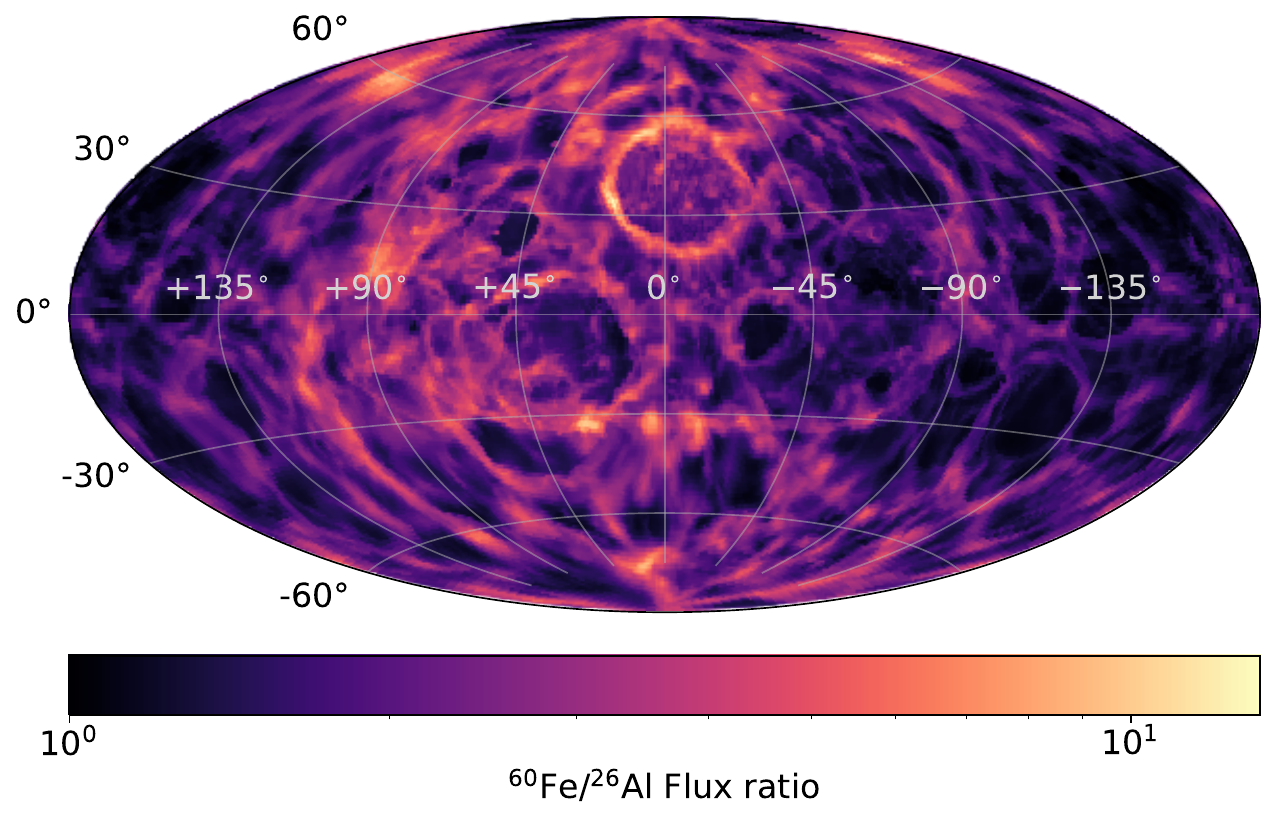}
    \caption{Flux ratio per pixel of the 1332\,keV line from \nuc{Fe}{60} vs. the 1809\,keV line from \nuc{Al}{26}. Left: Case 1. Right: Case 2. It is seen that the hot spots from \nuc{Al}{26} in Fig.\,\ref{fig:images_26Al} are turned into cold spots in these flux ratio maps. The reason is again the $3.8$ longer lifetime of \nuc{Fe}{60} which lets the $\gamma$-ray emission of this isotope appear more homogeneous. We note the scales of the flux ratio maps: in case 1, the contrast is particularly small, whereas in case 2, the strong accumulation in of \nuc{Fe}{60} already-mixed in the bubble walls increases the contrast up to a factor of $10$.}
    \label{fig:Fe60_vs_Al26_flux_ratio_comparison}
\end{figure*}

\subsubsection{Nuclear lines: 1809 and 1332\,keV}\label{sec:nuc_systematics}
In case 1, the nuclear lines from \nuc{Al}{26} and \nuc{Fe}{60} decay obtain their emissivities from geometrical and physical arguments, rather than hydrodynamics.
This means that every parameter, including the ejecta masses, sound velocity inside the bubble, positions of the supernovae, order of spherical harmonics expansion of the bubble walls, uptake efficiency of \nuc{Fe}{60} deposition on Earth, rotational velocity of the progenitors, and their metallicities, respectively, will change the appearance and total flux of the resulting images.
From the stellar parameters alone, the uncertainty in the ejecta masses is within a factor of $10$, which linearly impacts the total flux.
In our above calculations, we used optimistic, but not the upper bounds, ejecta yields, so that the the systematic uncertainties would be rather asymmetric with about 10\,\% in the positive and 90\,\% towards lower values.
However, an extrapolation towards super-solar metallicities could lead to even higher radioactive ejecta masses.
Comparing the yields and fluxes from case 1 to case 2 for the nuclear lines, the hydrodynamics simulations typically show a (much) larger present-day mass and therefore $\gamma$-ray line flux.
These appear particularly large because instead of two, 13--14 supernovae contribute to the \nuc{Al}{26} and \nuc{Fe}{60} mass in the hydrodynamics simulation.
Since the yields of these 13--14 supernovae are cumulating with time because the decay times of the radioisotopes are longer than the average delay time between successive supernovae ($\sim$0.7\,Myr), there is more nuclear material present in the Local Bubble.
This effect is particularly strong for \nuc{Fe}{60} with a lifetime of 3.8\,Myr and can readily explain the typical factors of $\sim$10 difference in mass and flux between cases 1 and 2.
The resulting fluxes from both cases can therefore be considered a bracket of systematic uncertainties for the 1809 and 1332\,keV line.
The isotropic contributions in both cases is similar, between 30--50\,\% for the 1809\,keV line and between 20--45\,\% for the 1332\,keV line, which is also reassuring that the geometrical and physical considerations of case 1 are reasonable.

\subsubsection{Positron annihilation line: 511\,keV}\label{sec:511_systematics}
In the case of positron annihilation, the systematics are similar to the nuclear lines:
While the total amount of (not annihilated) positrons at any given time is directly given by the \nuc{Al}{26} decay, Eqs.\,(\ref{eq:positron_luminosity_case1}) and (\ref{eq:total_positron_luminosity_case2}), that is, directly from the present mass of \nuc{Al}{26}, it is unknown which part of the positrons annihilates \emph{when}.
The annihilation regions are determined in case 1 from the simple argument of diffusion being inefficient for low-energy positrons, even though they may propagate path lengths of several kpc.
This means that the inner Local Bubble wall is the natural target, whose thickness can be estimated from weighing the diffusion time scale against the annihilation time scale, which results in thicknesses of $\sim$$10^{-3}$--$10^0$\,pc.
For case 1 we then assume that all positrons from \nuc{Al}{26} some time ago (according to the cooling time) annihilate inside a 1\,pc thick wall.
From the uncertainty of the \nuc{Al}{26} yields in case 1 alone, the total positron luminosity is uncertain by a factor of $\sim$10, ranging from $10^{36}$--$10^{37}\,\mathrm{e^+\,s^{-1}}$.
This directly impacts the 511\,keV flux, then ranging from $(2$--$20) \times 10^{-6}\,\mathrm{ph\,cm^{-2}\,s^{-1}}$.
The thickness of the walls also affects the measurable 511\,keV flux:
while thinner walls (in case 1) lead to the same flux, a distribution of the same number of positrons to thicker walls, e.g. 40\,pc as could be expected from theory, the flux may decrease due to more positrons at larger distances.
A similar argument can be considered for the 511\,keV emission using the hydrodynamics simulation, case 2:
here, the bubble walls are naturally about 20--30\,\% of the thickness of the bubble radius, which distributes much of the positrons further out and decreases the flux.
If then the bubble walls were thinner (assuming a larger boundary on the flow velocity), the flux may increase slightly by up to 10\,\% when only the high luminosity regions in Fig.\,\ref{fig:annihilation_luminosity_case2} would contribute.
Other annihilation channels, such as radiative recombination and direct annihilation with non-zero kinetic energy, do not significantly contribute in the scenarios discussed here.

\begin{figure*}[!ht]
    \centering
    \includegraphics[width=0.33\textwidth]{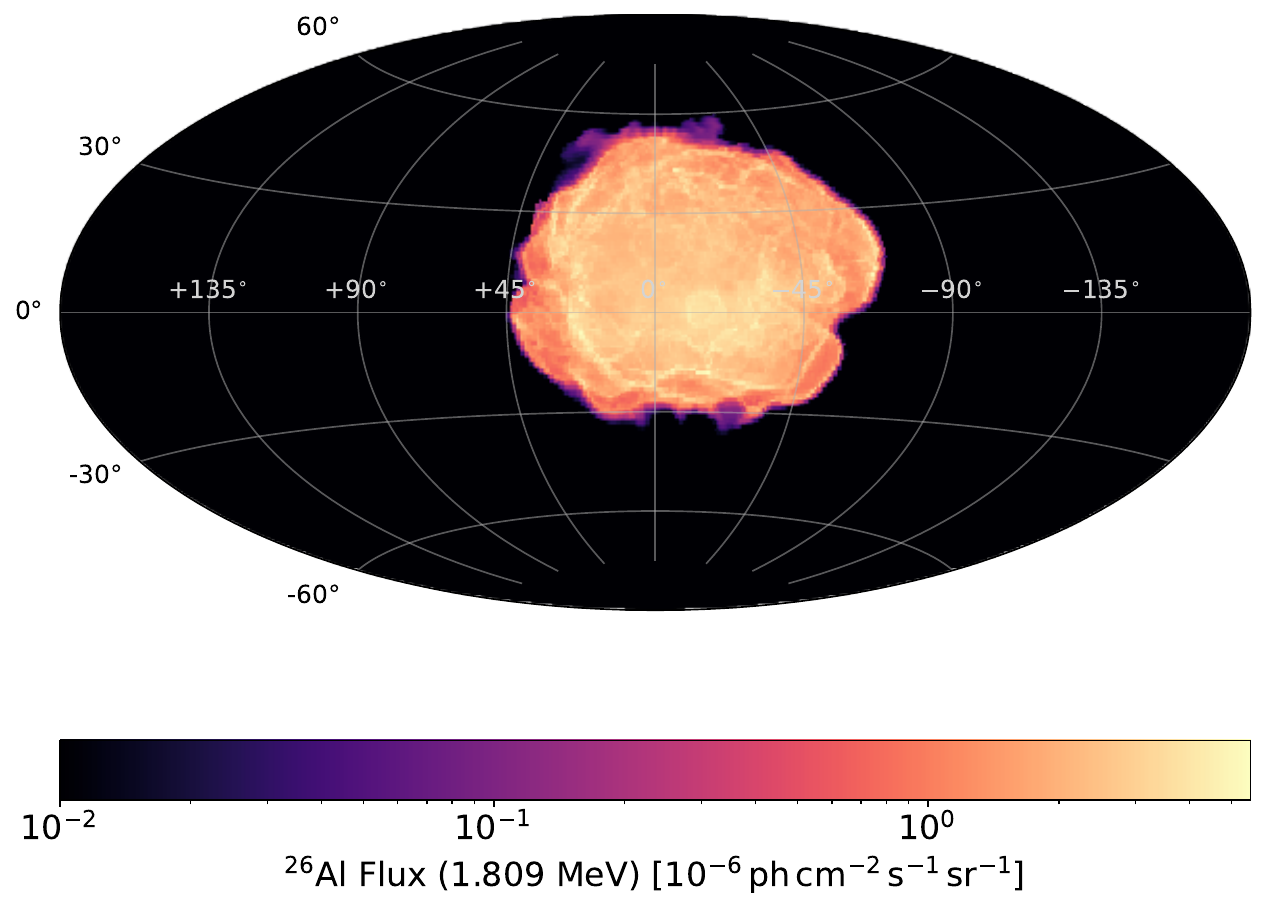}~
    \includegraphics[width=0.33\textwidth]{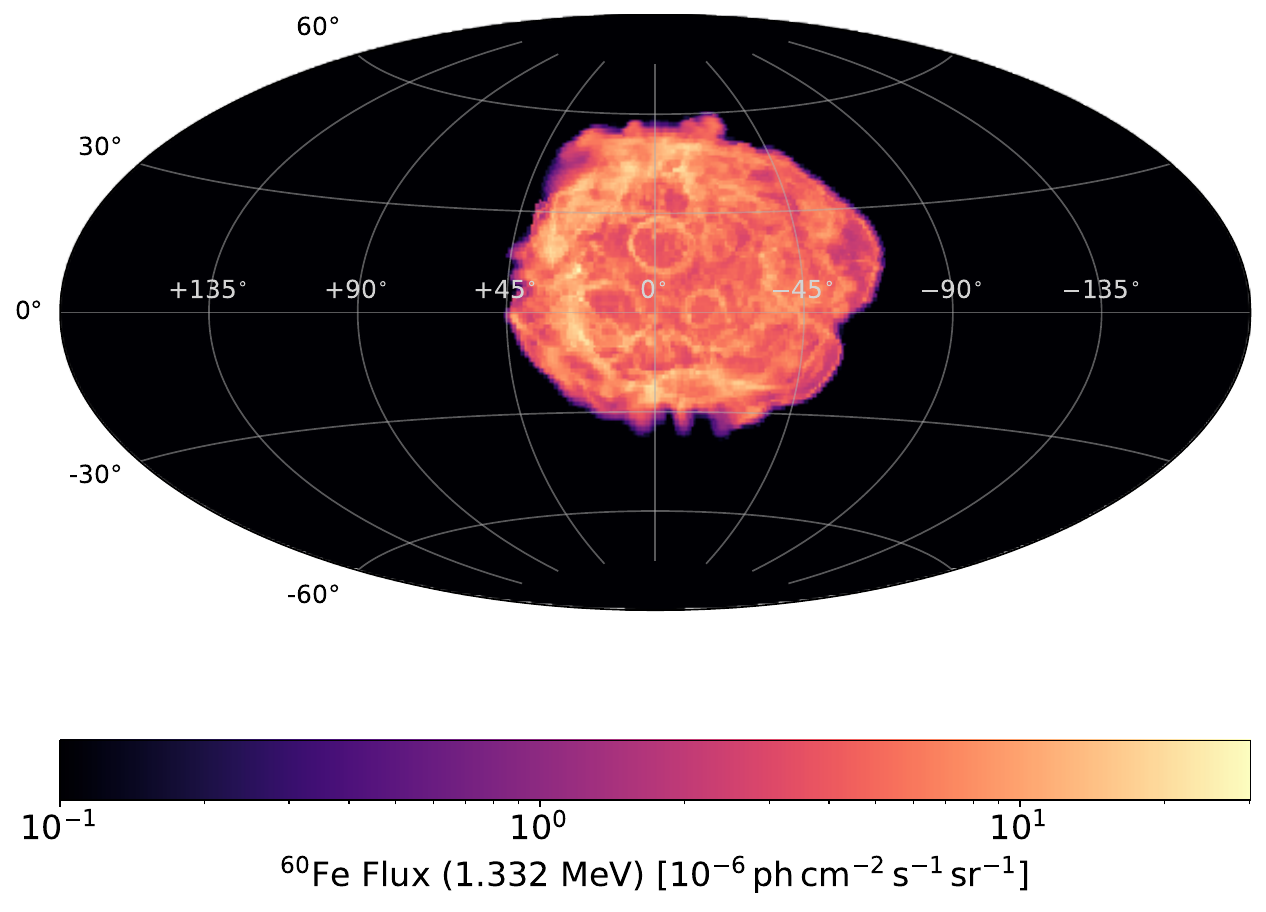}~
    \includegraphics[width=0.33\textwidth]{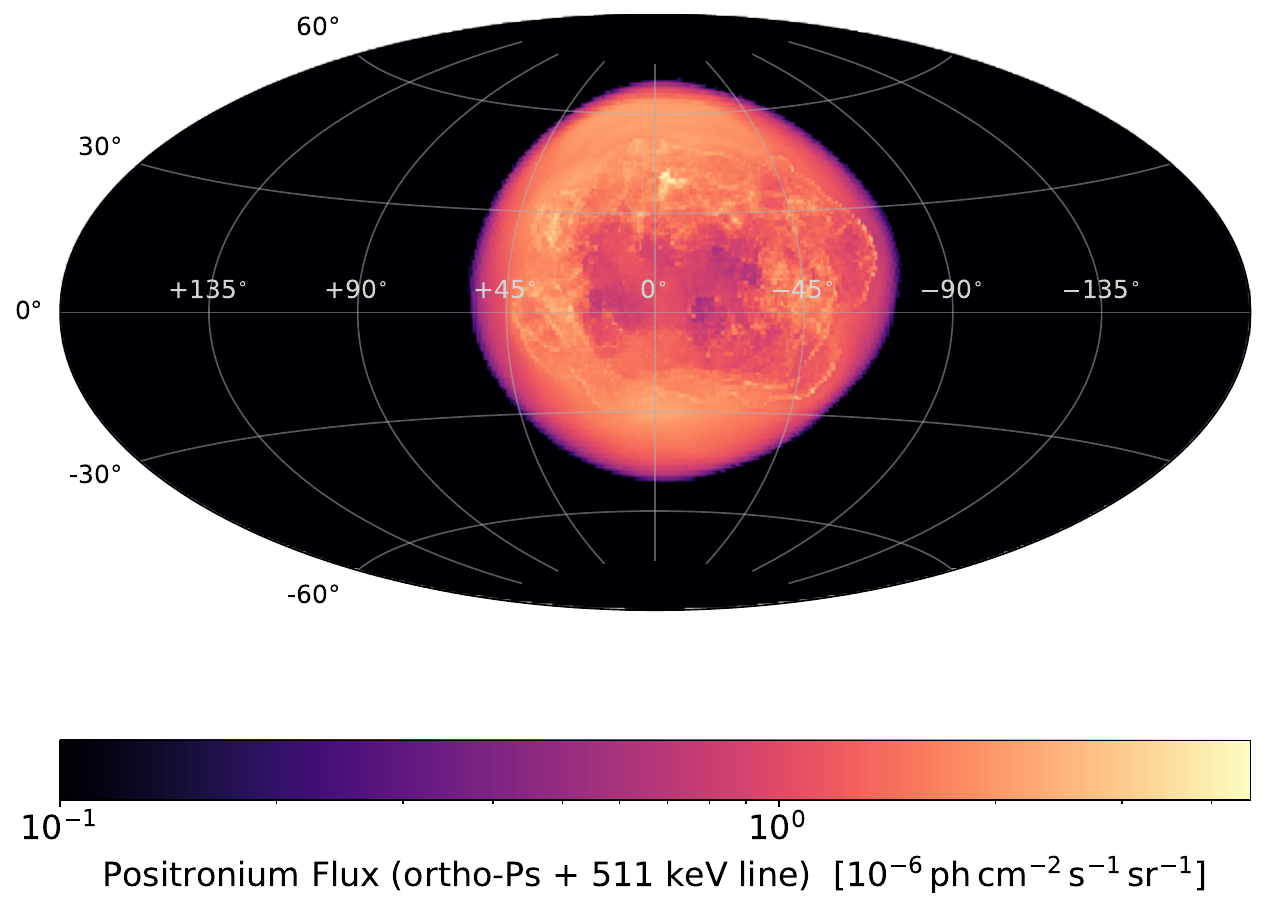}
    \caption{Appearance of the $\gamma$-ray emission from the Local Bubble at 1809\,keV (left), 1332\,keV (middle) and 511\,keV (right) if it were to be observed from outside (here: 200\,pc distance). It is evident that \nuc{Al}{26} and \nuc{Fe}{60} $\gamma$-ray emission is not co-spatial: the 1809\,keV line appears more homogeneous with slight enhancements at the bubble walls compared to the 1332\,keV line which shows an edge brightening. Positron annihilation only shows weak structures with bright spots from possible filaments. The thickness of the bubble wall is essentially determining the appearance of the 511\,keV with only a weak contrast from inside the bubble towards the bubble walls.}
    \label{fig:images_LB_outside_case2}
\end{figure*}

Two factors which may severely change the estimates here are the unsolved problem of propagation and the additional contribution from other positron sources.
If the propagation of low-energy positrons is rather ballistic, the positrons that are created inside the Local Bubble (or anywhere in the Galaxy) will not annihilate in their close vicinity in the bubble walls or molecular clouds.
This means that any isotropic contribution that we would measure at 511\,keV would be the cumulative effect of all different sources whose propagated positrons will annihilate in our vicinity \emph{by chance}.
In such as case, the isotropic contribution to the 511\,keV line may be even larger than what is discussed here.
The additional positron sources are novae, pulsars, accreting compact objects, and possibly dark matter, among others \citep[cf.][for a recent review]{Siegert2023_511}.
All of them may be found inside the Local Bubble and their positron production rate will add to the predictions from \nuc{Al}{26} alone and enhance the 511\,keV flux.
In summary, the 511\,keV flux estimated from the Local Bubble should be found at least at a level of $2 \times 10^{-6}\,\mathrm{ph\,cm^{-2}\,s^{-1}}$, with large uncertainties to the upper end with optimistic values of $(20$--$30) \times 10^{-6}\,\mathrm{ph\,cm^{-2}\,s^{-1}}$.

\subsection{\nuc{Fe}{60} to \nuc{Al}{26} ratio}\label{sec:Fe60_to_Al26_ratio}
For astronomical observations, the flux ratio of \nuc{Fe}{60} to \nuc{Al}{26}, $r_{60/26}$, is of high importance because it gauges the stellar evolution models for massive stars.
One has to distinguish between the mass ratio and the flux ratio of the two isotopes as they are only straight-forwardly related by a numerical factor if and only if the distributions of \nuc{Fe}{60} and \nuc{Al}{26} are identical.
We show in Fig.\,\ref{fig:Fe60_vs_Al26_flux_ratio_comparison} that this is not the case.
For both modelling assumptions, there is a certain contrast in the flux ratio maps which is the result of the different lifetimes of the isotopes again.
This contrast is further enhanced in the simulations because \nuc{Al}{26} is expelled by both winds of massive stars that live up to 0.88\,Myr before present and their supernovae, whereas \nuc{Fe}{60} is exclusively released by the supernovae, which leads to a slightly different geometry in case 1 and to more time available for mixing with the bubble walls in case 2.
The $\gamma$-ray line map from \nuc{Fe}{60} appears more homogeneous compared to \nuc{Al}{26} because in the latter, the last supernovae can still be identified, owing also to the cumulative effect of the wind contributions which only occur for \nuc{Al}{26}.
The flux ratio, then, also depends strongly on the used yields and the number of supernovae responsible for the total mass and therefore flux.
In case 1, the average flux ratio is about 0.32, whereas in case 2, it is 2.16.
It should be noted that the individual hotspots in Fig.\,\ref{fig:Fe60_vs_Al26_flux_ratio_comparison} (right, i.e. case 2) arise from turbulence which is not (or hardly) present in case 1.
It can, however, increase the flux ratio to values beyond 10.
Given the sensitivity of even COSI-SMEX, these pointed enhancements are probably too faint to be distinguishable and rather the average flux ratio will gauge the models.
Clearly, the more supernovae contribute, the higher is the flux ratio, so that a measurement of the ratio, which is certainly within reach for COSI-SMEX (see Sect.\,\ref{sec:COSI_sims}), will reveal how many recent supernovae were responsible for the \nuc{Fe}{60} anomalies measured on Earth (and the Moon).
We note that the individual stellar evolution models themselves also have an impact on the resulting flux ratio, but that the large difference in our models originate purely in the number of supernovae exploded, so that one can distinguish certain scenarios by finding $r_{60/26}$ larger or smaller than $1.0$ as a rough guideline.
This will also be in stark contrast to the Galactic-wide measurement of the flux ratio by \citet{Wang2020_Fe60} of $0.18 \pm 0.04_{\rm stat}$, and up to $0.4$ including systematic uncertainties.

\subsection{The Local Bubble viewed from the outside}\label{sec:outside_view_case2}
The Local Bubble simulation by \citet{Schulreich2023_LB} can be used as a surrogate model for (almost) any type of superbubble with an age of $\sim$$10$--$20$\,Myr.
We place the hydrodynamics simulation at a distance of 200\,pc from the Sun to perform line-of-sight integrations for \nuc{Al}{26}, \nuc{Fe}{60}, and positron annihilation, again, to inspect the appearance of a distant superbubble.
For the same setup, the fluxes are $4.7 \times 10^{-6}$, $14.5 \times 10^{-6}$, and $2.6 \times 10^{-6}\,\mathrm{ph\,cm^{-2}\,s^{-1}}$ for the 1809, 1332, and 511\,keV line, respectively.
Certainly, these values are subject to the same systematic uncertainties, independent of the observer position.

We show the appearance of the three $\gamma$-ray lines viewed from the `outside' in Fig.\,\ref{fig:images_LB_outside_case2}.
The most intriguing factor here is that, again, \nuc{Fe}{60} and \nuc{Al}{26} are not co-spatial.
This is important because this assumption is typically used in $\gamma$-ray data analyses, which require spatial templates.
Here, the emission at 1809\,keV appears more homogeneous with only slight enhancements from the turbulence, compared to the 1332\,keV line with edge brightenings.
The flux ratio, $r_{60/26}$, is then identical to the case from observing inside the bubble, and would, if this was observed by a $\gamma$-ray telescope, gauge the age of the bubble by population synthesis \citep[e.g.,][]{Siegert2023_PSYCO} rather than the number of supernovae that occurred.
The positron annihilation map (Fig.\,\ref{fig:images_LB_outside_case2}, right) appears less structured with possible filaments due to annihilation in cold and dense clumps inside the superbubble as a result of thermal and hydrodynamic instabilities (possibly molecular clouds) and bubble walls.
Again, the thickness is defined by the physical argument of where the transition region is from the bubble to the remaining interstellar medium of the Galaxy, not being neighboured by another bubble.
Positron annihilation shows only a weak contrast as a function of bubble radius, which is surprising giving the fact that positrons only annihilate in dense regions.
However, because of the thickness of the walls, the inhomogeneities wash out and leave the 511\,keV emission from a superbubble (at this age) as a distinct diffuse blob.
It may well be the case that positrons diffuse through the bubble walls into other, neighbouring superbubbles (e.g. Orion-Eridanus), so that the annihilation flux of one superbubble cannot be estimated strictly by only assuming the \nuc{Al}{26} ejecta masses.
Many positron populations from different bubbles would probably add together in their boundaries.

\section{Summary and outlook}\label{sec:conclusion}
In this work, we estimated the $\gamma$-ray line fluxes of the intermediate-lived radioactive isotopes \nuc{Fe}{60} and \nuc{Al}{26}, and the subsequent annihilation of positrons from the $\beta^+$-decay of \nuc{Al}{26}, in the Local Bubble.
We based our estimates on two assumptions, one being a geometrical model that was derived to match the 3D dust extinction maps \citep{Pelgrims2020_LocalBubble}, and one being a hydrodynamics simulation that was used to explain the deep-sea \nuc{Fe}{60} deposits on Earth, among others \citep{Schulreich2023_LB}.
Despite these completely different approaches, the resulting fluxes for the 1809\,keV line of \nuc{Al}{26}, the two lines at 1173 and 1332\,keV of \nuc{Fe}{60}, and the positron annihilation line after the formation of Positronium at 511\,keV, are consistent within an order of magnitude.
The fluxes range from $(3$--$20) \times 10^{-6}\,\mathrm{ph\,cm^{-2}\,s^{-1}}$ for \nuc{Al}{26}, $(4$--$40) \times 10^{-6}\,\mathrm{ph\,cm^{-2}\,s^{-1}}$ for \nuc{Fe}{60}, and $(2$--$20) \times 10^{-6}\,\mathrm{ph\,cm^{-2}\,s^{-1}}$ for positron annihilation.
The fundamental difference in the assumptions is the number of supernovae within the last $\sim$10\,Myr that lead to the diffuse $\gamma$-ray flux now.
While our geometrical model assumes only two supernovae, the hydrodynamics simulation assumes 13--14 supernovae.
This factor of 6--7 is directly reflected in the flux of the 1809\,keV line, and leads to even larger differences in the case of \nuc{Fe}{60}, up to a factor of 10.
By measuring these fluxes inside the Local Bubble against the Milky Way background, one could further constrain models of the formation of the Local Bubble and the number of supernovae responsible for its current size and shape.
In fact, with a 100-fold increase in sensitivity with respect to current telescopes \citep[that is, INTEGRAL/SPI;][]{Winkler2003_INTEGRAL,Vedrenne2003_SPI}, one could even determine the direction of the last, that is, most recent, supernova in the Local Bubble.

While current $\gamma$-ray telescopes cannot measure these almost-isotropic signals, the future COSI-SMEX \citep{Tomsick2023_COSI} satellite mission, slated for launch in 2027, will have a sensitivity that allows estimates of these $\gamma$-ray line fluxes.
The time to reach a $5\sigma$ signal in the two nuclear lines varies between 0.1 and 4\,yr against the instrumental background and the Milky Way background.
We emphasise that these numbers should be considered with caution as the identification of large-scale signals is not trivial, and that the estimates for the instrumental background and exposure are subject to small changes until the launch.
In addition, there will be contributions from the Galactic diffuse continuum \citep[e.g.,][]{Siegert2022_MWdiffuse} that will contribute to the line flux in small energy bands.
Thanks to the excellent spectral resolution of COSI-SMEX, these contributions will be marginal in the case of 511 and 1809\,keV, but can lead to a considerable bias in the case of the \nuc{Fe}{60} lines at 1173 and 1332\,keV.
If it is possible to determine the Cosmic Gamma-ray Background \citep[e.g.,][]{Inoue2014_CGB} in the MeV range with COSI with high spectral resolution, one could compare the isotropic fractions of these lines attributed to the Local Bubble:
We find that the ubiquitous contribution of the \nuc{Al}{26} and \nuc{Fe}{60} lines is between 20 and 50\,\% of the total Local Bubble flux, thus also on the order of $10^{-6}\,\mathrm{ph\,cm^{-2}\,s^{-1}}$.

The assumptions used in this study to estimate the 511\,keV line flux from the \nuc{Al}{26} decay in the Local Bubble walls results in either a very strong contribution on the order of $10^{-5}\,\mathrm{ph\,cm^{-2}\,s^{-1}}$ with an isotropic contribution of $20$--$40$\,\% if based on our geometrical model, or rather weak fluxes around $10^{-6}\,\mathrm{ph\,cm^{-2}\,s^{-1}}$ with $10$--$15$\,\% isotropic contribution if based on the hydrodynamics simulation.
This largest difference in our estimates probably results from the unknown propagation properties of low-energy positrons:
While in the geometrical model, we find that, based on a uniform diffusion coefficient of $10^{28}\,\mathrm{cm^2\,s^{-1}}$, all positrons naturally annihilate in the nearest $0.1$--$1.0$\,pc of the inner Local Bubble wall, the hydrodynamics simulation provides 3D distributions of hydrogen densities and temperatures for the positrons to annihilate in with specific cross sections.
We note that these are two extremely different assumptions, but which could be tested with a next generation telescope like COSI.

By investigating the appearances of the \nuc{Al}{26} and \nuc{Fe}{60} maps of only the Local Bubble, the most important finding in this study is that the emission at 1809\,keV and 1332\,keV is not co-spatial.
This is independent of the point of the observer and the modelling assumptions.
If the Local Bubble is viewed from the `outside', the \nuc{Al}{26} emission at 1809\,keV appears rather homogeneous with a slightly brighter central region compared to the outskirts with indication of turbulence.
In contrast, the \nuc{Fe}{60} emission at 1173 and 1332\,keV is edge-brightened because the \nuc{Fe}{60} had more time to mix with the denser hydrogen gas of the bubble wall until it is decaying with a 3.8 longer lifetime as \nuc{Al}{26}.
The wind contributions from \nuc{Al}{26} furthermore lead to a quasi-steady flow which shapes the $\gamma$-ray image in a different way compared to \nuc{Fe}{60} whose nucleosynthesis input is always explosive.
In the context of $\gamma$-ray data analysis, which relies predominantly on spatial templates to determine the spectra and therefore fluxes of emission regions, the consensus assumptions that \nuc{Al}{26} and \nuc{Fe}{60} emission is co-spatial must be revised.
If and only if the emission regions are identical can the flux ratios be converted into mass and production ratios and therefore inform about stellar evolution models of massive stars.

\begin{acknowledgements}
Saurabh Mittal and Hiroki Yoneda acknowledge support by the Bundesministerium f\"ur Wirtschaft und Energie via the Deutsches Zentrum f\"ur Luft- und Raumfahrt (DLR) under contract number 50 OO 2219.
\end{acknowledgements}

\bibliographystyle{aa} 
\bibliography{thomas,schulreich} 

\appendix
\section{Derivation of Eq.\,(\ref{eq:general_solution_deq})}\label{app:diff_eq_solution}
Starting from the steady state assumption of Eq.\,(\ref{eq:diff_ann_equation}),
\begin{equation}
    \frac{1}{n^*(r)}\,\frac{\partial n^*(r)}{\partial t} = 0\,,
\end{equation}
the equation to solve is
\begin{equation}
\begin{split}
    0 &= 4\,\pi\,\frac{D}{\kappa\,r}\,\left[n^*(r)\,r^2 - n^*(r+\mrm{d}r)\,(r + \mrm{d}r)^2 \right] \\
    &\quad-4\,\pi\,r^2\,\dot{a}(T)\,n_{\rm H}(r)\,n^*(r)\,\mrm{d}r\,.
\end{split}
\end{equation}
Expanding the terms gives
\begin{equation}
\begin{split}
    0 &= \frac{D}{\kappa r}\,\left[n^*(r)\,r^2 - n^*(r + \mrm{d}r)\,r^2 - 2\,r\,n^*(r + \mrm{d}r)\,\mrm{d}r\right.\\
    &\quad\left.-n^*(r + \mrm{d}r)\,\mrm{d}r^2\right] - r^2\,\dot{a}(T)\,n_{\rm H}(r)\,n^*(r)\,\mrm{d}r\,,
\end{split}
\end{equation}
of which the last term in the parenthesis approaches zero.
Collecting the terms with powers of $r$ results in
\begin{equation}
\begin{split}
    0 &= \frac{D}{\kappa}\,\left[ n^*(r)\,r - n^*(r + \mrm{d}r)\,r - 2\,n^*(r + \mrm{d}r)\,\mrm{d}r \right]\\
    &\quad-\dot{a}(T)\,n_{\rm H}(r)\,n^*(r)\,r^2\,\mrm{d}r\,,
\end{split}
\end{equation}
where a spatial derivative of $n^*(r)$ with respect to $r$ can be identified, so that
\begin{equation}
\begin{split}
    0 &= \frac{D}{\kappa}\,\left[ -r\,\frac{\mrm{d}n^*(r)}{\mrm{d}r}\,\mrm{d}r - 2n^*(r + \mrm{d}r)\,\mrm{d}r \right]\\
    &\quad-\dot{a}(T)\,n_{\rm H}(r)\,n^*(r)\,r^2\,\mrm{d}r\,,
\end{split}
\end{equation}
which can be reduced to
\begin{equation}
    0 = -\frac{\mrm{d}n^*(r)}{\mrm{d}r} - \frac{2}{r}\,n^*(r + \mrm{d}r) - \frac{\kappa}{D}\,\dot{a}(T)\,n_{\rm H}(r)\,r\,n^*(r)\,.
\end{equation}
In the limit of a thin shell slice, $\mrm{d}r \rightarrow 0$, the new differential equation reads
\begin{equation}
    0 = -\frac{\mrm{d}n^*(r)}{n^*(r)} - \frac{2}{r}\,\mrm{d}r - \frac{\kappa\,\dot{a}(T)\,n_{\rm H}(r)}{D}\,r\, \mrm{d}r\,.
\end{equation}
By simple integration, we arrive at the solution
\begin{equation}
    n^*(r) = n_0^*\,\exp\left\{- \int \left[ \frac{2}{r} + \frac{\kappa\,\dot{a}(T)\,n_{\rm H}(r)}{D}\,r \right]\,\mrm{d}r \right\}\,,
\end{equation}
which is equivalent to Eq.~(\ref{eq:general_solution_deq}).

\end{document}